\documentclass[12pt,preprint,usenatbib]{emulateapj}

\usepackage{amssymb}
\usepackage{times}
\usepackage{pifont} 
\usepackage{rotating}
\usepackage{longtable}

\newcommand{\pasa}{PASA}
\newcommand{\teff}{$T_{\mathrm{eff}}$}
\newcommand{\muhz}{$\mu$Hz}
\newcommand{\numax}{$\nu_{\mathrm{max}}$}
\newcommand{\dnu}{$\Delta\nu$}

\newcommand{\msol}{M$_\odot$}

\newcommand{\galaxia}{\textit{Galaxia}}
\newcommand{\kepler}{\textit{Kepler}}

\shorttitle{K2 GAP DR-1}
\shortauthors{Stello et al.}

\begin{document}

\title{The K2 Galactic Archaeology Program Data Release 1: asteroseismic results
  from Campaign 1}

\author{
Dennis~Stello\altaffilmark{1,2,3}, 
Joel~Zinn\altaffilmark{4},  
Yvonne~Elsworth\altaffilmark{5,2},
Rafael~A.~Garcia\altaffilmark{6},
Thomas~Kallinger\altaffilmark{7},
Savita~Mathur\altaffilmark{8},
Benoit~Mosser\altaffilmark{9},
Sanjib~Sharma\altaffilmark{2},
William~J.~Chaplin\altaffilmark{5,2},
Guy~Davies\altaffilmark{5,2},
Daniel~Huber\altaffilmark{2,3,10}, 
Caitlin~D.~Jones\altaffilmark{5,2},
Andrea~Miglio\altaffilmark{5,2},
Victor~Silva~Aguirre\altaffilmark{3}
}
\altaffiltext{1}{School of Physics, University of New South Wales, NSW 2052, Australia}
\altaffiltext{2}{Sydney Institute for Astronomy (SIfA), School of Physics, University of Sydney, NSW 2006, Australia}
\altaffiltext{3}{Stellar Astrophysics Centre, Department of Physics and Astronomy, Aarhus University, Ny Munkegade 120, DK-8000 Aarhus C, Denmark}
\altaffiltext{4}{Department of Astronomy, The Ohio State University, Columbus, OH 43210, USA}
\altaffiltext{5}{School of Physics \& Astronomy, University of Birmingham, Edgbaston, Birmingham, B15 2TT, UK}
\altaffiltext{6}{Laboratoire AIM, CEA/DSM -- CNRS - Univ. Paris Diderot -- IRFU/SAp, Centre de Saclay, 91191 Gif-sur-Yvette Cedex, France}
\altaffiltext{7}{Institute of Astrophysics, University of Vienna, Türkenschanzstrasse 17, Vienna 1180, Austria}
\altaffiltext{8}{Center for Extrasolar Planetary Systems, Space Science Institute,  4750 Walnut street Suite 205 Boulder, CO 80301 USA}
\altaffiltext{9}{LESIA, Observatoire de Paris, PSL Research University, CNRS, Universit\'e Pierre et Marie Curie, Universit\'e Paris Diderot, 92195 Meudon, France cedex, France}
\altaffiltext{10}{SETI Institute, 189 Bernardo Avenue, Mountain View, CA 94043, USA}

\begin{abstract}
NASA's K2 mission is observing tens of thousands of stars along the
ecliptic, providing data suitable for large scale asteroseismic analyses to
inform galactic archaeology studies. Its first campaign covered a field
near the north galactic cap, a region never covered before by large
asteroseismic-ensemble investigations, and was therefore of particular interest 
for exploring this part of our Galaxy.  
Here we report the asteroseismic analysis of all stars selected by the K2
Galactic Archaeology Program during the mission's ``North Galactic Cap''
campaign 1.  Our consolidated analysis uses six independent methods to
measure the global seismic properties, in particular the large frequency
separation, and the frequency of maximum power.  
From the full target sample of 8630 stars we find about 1200 oscillating
red giants, a number 
comparable with estimates from
galactic synthesis modeling.  Thus, as a valuable by-product we find roughly 7500
stars to be dwarfs, which provide a sample well suited for galactic
exoplanet occurrence studies because they originate from our simple and easily 
reproducible selection function.  
In addition, to facilitate the full potential of the data set for galactic
archaeology we assess the detection completeness of our sample of
oscillating red giants.  We find the sample is at least near complete for
stars with $40 \lesssim $ \numax/\muhz\ $\lesssim 270$, and 
\numax$_\mathrm{,detect} < 2.6\times 10^6 \cdot 2^{-\mathit{Kp}}\,$\muhz. 
There is a detection bias against helium core burning stars with
\numax\ $\sim 30\,$\muhz, affecting the number of measurements of \dnu\ and
possibly also \numax.  Although we can detect oscillations down to
$\mathit{Kp}=15$, our campaign 1 sample lacks enough faint giants to assess
the detection completeness for stars fainter than $\mathit{Kp}\sim 14.5$. 

\end{abstract}

\keywords{stars: fundamental parameters --- stars: oscillations --- stars:
  interiors --- planetary systems}


\section{Introduction} 
From the birth of \kepler's second-life mission, K2 \citep{Howell14}, grew
the K2 Galactic Archaeology Program (GAP) \citep{Stello15}.  With this program
we aim to measure convectively-driven, or solar-like, oscillations in tens of
thousands of red giants using the high-precision photometry from K2.  
The asteroseismic imprint in the data combined with ground-based
measurements of effective temperature and metallicity will allow us to infer
stellar properties such as radius and mass, and hence distance and age,
even for stars out to a few tens of kilo parsec \citep{Mathur16}.  This in
turn will enable us to probe the structure and evolution of vast regions of
the Galaxy not probed before by asteroseismic means \citep[e.g.][]{Miglio12}.    
It has become clear that the power of such seismic-inferred galactic
studies, depends strongly on how well we understand the target
selection and the detection biases involved \citep{Sharma16}. 
The K2 GAP target selection has been particularly designed to gain full control
of the selection bias, and hence to mitigate some of the problems encountered
using stars from CoRoT or the original \kepler\ mission. 
However, for the program to reach its full potential we also need to be able to 
predict any detection bias in the asteroseismology in order to make robust
inferences based on galactic model comparisons.  

In this paper, we present the seismic analysis of all stars selected
by the K2 GAP in campaign 1 (C1), and release, to the community, the full
seismic data set of the large frequency separation, \dnu, and the frequency
of maximum power, \numax.  After an initial description of the target
selection (Sect.~\ref{targetselect}), we give a brief overview of the methods
used to measure the seismic properties (Sect.\ref{methods}), before
presenting the results in Sect.~\ref{results}.
In Sect.~\ref{detectbias} we assess the detection bias and its effect on
the sample completeness before concluding in Sect.~\ref{conclusion}.

\section{Campaign-1 target selection and observations}\label{targetselect}
Our prime targets are red giants. The advantage of red giants is that they
are intrinsically bright and show larger amplitude oscillations
compared to dwarfs. Oscillations can therefore be detected even for distant
stars.  Most giants also oscillate with frequencies that are well sampled
by K2's main, long cadence, observation mode. This enables large numbers
of stars to be observed as required to perform robust stellar population studies. 

\subsection{Target selection strategy}
\begin{figure*}
\includegraphics[width=17.6cm]{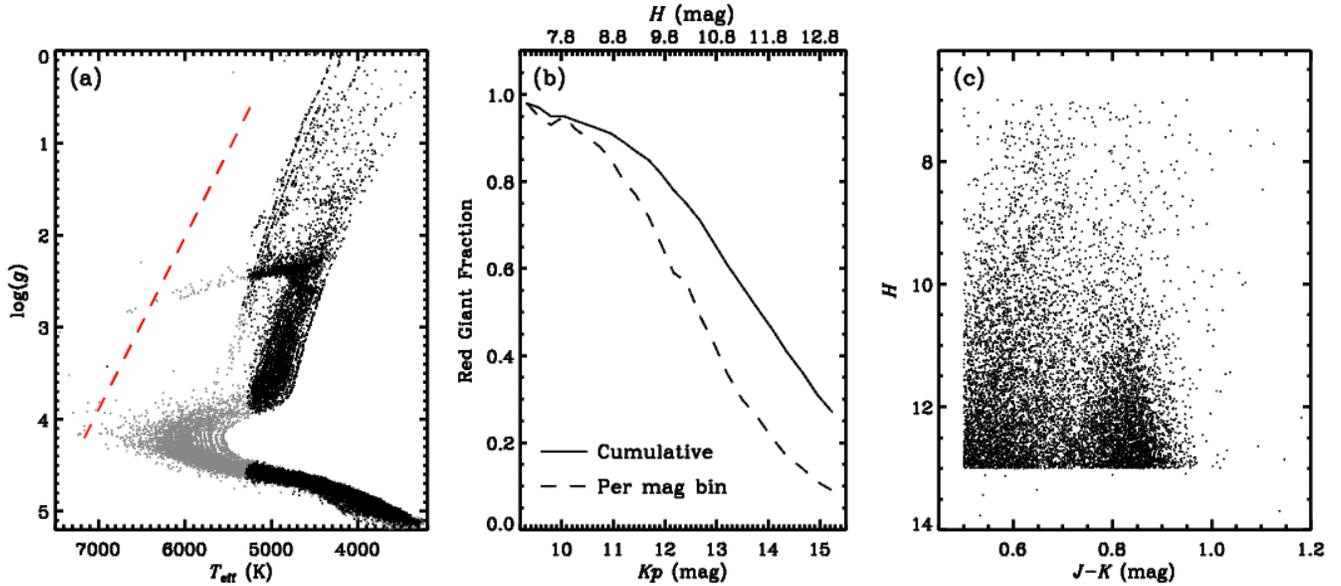}
\caption{(a) Kiel diagram of a synthesis population of the K2 C1
  stars based on a \galaxia\ simulation (Sect.~\ref{synergy}). Black
  symbols show stars that fulfill the K2 GAP selection criteria. The
  dashed red line indicate the approximate location of the red edge of the
  classical instability strip. 
         (b) Estimated giant fraction of K2 GAP targets per magnitude bin
  (dashed) and cumulative going form bright to faint (solid), 
  based on a \galaxia\ simulation. The approximate $H$-band magnitude
  is shown on the top axis. 
         (c) Color-magnitude diagram of stars observed by K2 during C1 that
   were selected from the K2 GAP target list.
\label{cmd}} 
\end{figure*} 
We deliberately follow a simple color-magnitude target selection 
to ensure reproducibility and to strengthen the synergy with the selection
criteria of large-scale high-resolution spectroscopic surveys from ground
such as APOGEE \citep{Majewski10}, and Galah \citep{DeSilva15}.  
Specifically, for C1 (proposal
GO1059)\footnote{http://keplerscience.arc.nasa.gov/index.shtml} we
selected all stars with $J-K>0.5$ and $7<H<13$ 
using 2MASS, and sorted our target list by $H$-band 
magnitude \citep{Skrutskie06}\footnote{Stars were selected only if 2MASS
  bflag=1, cflag=0, xflg=0, aflg=0, $\mathrm{prox}>6.0$, and photometry quality flags
  (qflg) equal to A or B. See 2MASS website, vizier 2MASS table, or 
  the APOGEE target selection paper by \citet{Zasowski13} for
  more information on flags.}.  
Stars with $J-K>0.5$ already observed by APOGEE (258) and RAVE
(another 147) at the time of target selection were bumped up as top
priority targets, and for those we allowed the eight stars fainter than
$H=13$ to enter our list. 
Figure~\ref{cmd}a shows $\log g$ versus \teff\ for a synthesized population
of stars representative of the K2 C1 field, which we generated 
using the \galaxia\ Milky Way synthesis tool \citep{Sharma11}. We have
highlighted the stars that follow the K2 GAP selection function. It
illustrates how our color-cut is the sweet spot to ensure we do not loose
many helium-core burning red giants while not including 
too many main sequence and subgiant stars for which we can not
detect oscillations with K2's long cadence mode. Figure~\ref{cmd}b
shows the expected giant fraction based on the \galaxia \ simulation, and
Figure~\ref{cmd}c shows the color-magnitude diagram of the K2-GAP-selected
stars that were observed by K2.  

Of the total 21,648 C1 targets observed by K2, 8,630 are on the K2 GAP
target list.  This number is further broken down into 8,385 stars
explicitly sourced from our target list, hence following our 
selection approach, and 245 serendipitous stars sourced from other K2
observing programs, which are therefore following a different, and rather
complex, selection function. 
We further note that there were 155 stars, randomly distributed in priority, 
among our top 8,540 ranked targets, which were not observed, hence
resulting in the 8,385 observed K2 GAP-selected stars in total.  The
missing 155 stars represent an unbiased set, which were deselected by the
mission because they, or a saturated neighbor, were deemed too close to an
edge of active silicon on the detector. 
Hence, they do not cause concern in terms of biasing the selection function.
In summary, our color-magnitude selection can be regarded complete down to
$H=12.9$, corresponding to 14.5-16.5 in the \kepler\ band-pass,
$\mathit{Kp}$, depending on color. 

All K2 GAP targets were observed in the spacecraft's 29.4 minute long
cadence mode for about 80 days from 2014 May 30 to 2014 Aug 21.  We adopted
the light curves from \citet{VanderburgJohnson14} and \citet{Vanderburg15}, which we
high-pass filtered and gap-filled following the approach by \citet{Stello15}.
This approach should minimise the artifacts in the data caused by the
near-regular firing of the on-board thrusters used to stabilise the spacecraft
role angle every 6-hrs. 


\subsection{Synergy with exoplanet research}\label{synergy}
Our color-magnitude selection is expected to include a significant
fraction of dwarfs and subgiants, particularly for high
Galactic latitude fields like C1 (Figure~\ref{cmd}b).
Due to the simplicity of the selection function, these dwarfs and subgiants will
form a unique subset to study exoplanet occurrence rates as a function of
different stellar populations in our galaxy. The giant-dwarf
classifications provided by our program will also be valuable to prevent
biases when calculating planet occurrence rates, which sensitively depend
on stellar classifications \citep{Burke15}.


Our subsequent seismic analysis will show oscillations only in 
stars with roughly $1.9 < \log g < 3.2$ \citep{Stello15}, but for stars with
$\log g<1.9$ we can expect to see the signature of granulation, which
will reveal that they are intrinsically bright giants (Sect.~\ref{methods}).
Hence, the stars with no detected oscillations or granulation will most likely
be the dwarfs and subgiants that could be used for exoplanet occurrence
rate studies.
From a population synthesis model of our C1 selection using
\galaxia\ \citep{Sharma11} we expect our sample to include about 7500 stars
with $\log g > 3.2$, almost all dwarfs ($\log g > 4.2$) and only about 100
subgiant and low luminosity red giant branch stars ($3.2<\log g<4.2$) (see Sect.~\ref{results}). 



\section{Approach and methods}\label{methods}
The range of oscillation frequencies detectable from a single campaign of
K2 data is set by the length and the sampling rate of the time series,
which defines the lower and upper frequency limits, respectively.  These
correspond to a range in \numax\ of $\sim 10$-$270\,$\muhz\
\citep{Stello15}, 
or roughly $1.9<\log g<3.2$. 
Because our selection approach includes stars outside this range,
and because it is critical to understand the seismic detection bias for
galactic archaeology purposes, it becomes particularly important that we
assess robustly if a star is a giant, with $10<$ \numax/\muhz\ $<270$, or not.  

We receive 4,000-10,000 new K2 GAP targets for
each campaign.  With this rate of new stars to be analysed, and the need to
robustly determine which of them are giants, calls for a new
approach to analyse and consolidate the seismic
results compared to previous analyses based on the \kepler\ field. 
Those analyses were focusing on providing measurements of stars that
where clearly showing oscillations, assessed by the level of agreement
between results from independent methods \citep{Hekker11,Hekker12,Pinsonneault14}.  
This approach is prone to favour/select stars with the
clearest detections rather than selecting all stars, introducing a bias
that is difficult to mimic in subsequent comparisons with galactic
synthesis models. 

Here we use six independent analysis pipelines (called CAN, COR,
BHM, A2Z, SYD, and BAM)\footnote{We refer to the pipelines with their
  commonly used nicknames.}  
to analyse the light curves in order to
extract the large frequency separation, \dnu, and the frequency of maximum
power, \numax\ \citep[see e.g.][~for more discussion on those two
observables]{KjeldsenBedding95,ChaplinMiglio13}. 
Except for BHM, we also measure the amplitude of the 
oscillations, or some equivalent measure of oscillation strength for
validation purposes. 
Pipeline leads were asked to only return results that they trusted, because
comparisons between pipelines would not be used to remove outliers.
Instead, we rely on the combined set of results from all pipelines to
generate a list of stars for which at least one pipeline was able to detect
oscillations (meaning at least measure \numax), and for that set to make
careful visual inspections  
to verify if the stars indeed show signs of oscillations or granulation
in the power spectrum. 


\subsection{Pipelines}\label{pipelines}
In the following we describe briefly each pipeline and how some of them
used post processing criteria to select the final sample of stars believed
to show oscillations.  It should be stressed that only three pipelines
(CAN, BHM, and BAM)
determine the likelihood of a star showing oscillations based purely on the
power and time scale of the seismic (and granulation) signal.  The others
require a clear presence of a regular frequency pattern, from which \dnu\
can be measured.  The latter is a more strict criterion and hence leads to
fewer detections of oscillating stars.

{\bf CAN} uses the following Bayesian nested-sampling-based scheme for the
fitting and estimation of the measurement uncertainties. 
First, the `typical' time scale in the time series is found by fitting a
sinc function to the auto-correlation function (ACF) of the time series
\citep{Kallinger16}, and a $\chi^2$ goodness of fit test is carried out.
Then, it is evaluated if the rms of the time series is within a factor of
five from the expected value given the ACF timescale, which makes use of an
empirical rms-time scale relation (analogous to using an amplitude-\numax\
relation).  If a star passes both the goodness of fit test and fall within
the rms-time scale relation, it is accepted as a `red giant candidate', for
which the following, and most restrictive, step is performed.
By fitting the power spectrum with and without a Gaussian (to account
for an oscillation power excess) on top of the background, the Bayesian model
evidence determines if the power excess is statistically significant or not.
Here, the range in \numax\ for the location of the Gaussian is guided by
the ACF time scale \citep{Kallinger16}. 
If a star is deemed to not show any significant power excess implies that the
\numax\ from the fit cannot be trusted and the star is deemed a
non-detection.  To tension the number of accepted false positive against
the number of reject true positives, the applied significance threshold
was informed by visual inspection of the power spectra that felt close to
the acceptance/rejection cut-off.  This approach returned 1106 oscillating stars.
The central three radial orders of the power excess is then fitted by a
sequence of Lorentzian profiles representing oscillation modes and 
parameterised by the frequency of the central radial mode and a large and
small frequency separation \citep{Kallinger10a,Kallinger14}.
However, the fit to extract \dnu\ returned
robust results only for a subset of the stars (583 stars).

{\bf COR}: The method is the same as used for CoRoT or \kepler\ stars
\citep{MosserAppourchaux09}. The large separation is detected first, from the 
autocorrelation of the time series computed as the Fourier spectrum of the
filtered Fourier spectrum of the signal (EACF). A simple statistical test
based on the H0 hypothesis is used to assess the significance of the \dnu\
detection from the value of the EACF. 
In case of positive detection, the 
other global seismic parameters are determined. A star is regarded to 
show solar-like oscillations if its \dnu, FWHM of the smoothed power
excess, mean height of the power excess at \numax, and height of the
background at \numax\ all closely follow the empirically calibrated mean
relations as a function of \numax\ \citep{Mosser12a}.
For K2, a special treatment is applied to cope with the spacecraft thruster
firing frequency. It consists in using the empirical scaling relations with
stronger constraints when \numax\ is found to be near the firing frequency
or its harmonics. Uncertainties are calibrated functions of the EACF
parameter \citep[see][~for details]{MosserAppourchaux09}.  COR returned
results for all stars thought to provide robust \dnu\ and \numax\ values
(778 stars). 

{\bf BHM} uses a three-stage approach, described in detail in Elsworth (in
prep.).  To summarise:  First, the white-noise-subtracted power spectrum is
divided by a heavily smoothed version of the raw power spectrum,
representing the noise profile, to create the signal-to-noise ratio (SNR)
as function of frequency.  The likelihood of each frequency bin
representing just noise is then calculated according the H0
hypothesis. Next, we apply the H1 hypothesis to a series of overlapping
frequency segments of gradually increasing central frequency along the
SNR-spectrum. The width of each segment changes as function of its central
frequency following the empirical oscillation power width-\numax\ relation
from \citet{Mosser12a}.  The combination of H0 and H1 is used to form an
odds ratio of there being oscillation signal in a given segment.  If a
segment shows detection of oscillation power, the next two stages provide
estimates of \numax\ and \dnu.  This part of the method follows largely the
OCT method described in \citet{Hekker10}, but with some extra selection
criteria needed because of the short duration of the time series. For \dnu,
stringent tests are applied to mitigate against the measured \dnu\ being
corrupted by the frequency spacing between the radial and octupole
modes. These tests meant that for many of the stars only the \numax\ values
is considered reliable. In total 1031 stars have a robust measurement of
\numax, while only 198 have also a \dnu\ value.  

{\bf A2Z+} is an improved version of the A2Z pipeline described by
\citet{Mathur10}, but we refer to it in
this paper simply as A2Z.  It uses the same method as COR to determine a
first estimate of \dnu, and the frequency range where that signal is
present, from which a collapsed echelle diagram is formed to determine the
location of the radial modes. \numax\ is determined by fitting a Gaussian
in top of the background to the power spectrum.  The guess for granulation
timescale used in the fit is based on the results from \citet{Mathur11}.
For stars where the highest radial mode was statistically significant at
the 99\% level, \dnu\ is recomputed from the power spectrum of the power
spectrum (PS2) where the dipole modes are masked out. Here, we use only the 
central orders of the spectrum centered on the highest radial mode.
The number of radial orders used is typically four but depends slightly on the
signal-to-noise ratio and \numax.
The power spectrum and echelle diagram are used to visually verify the
\dnu\ measurements, and only stars with convincing detections are kept. 
Finally, as a post-processing criteria, stars are kept only if the slope of
a linear fit to the power spectrum in log-log space is steeper than
-0.5. In order words, if the granulation signal is not clear the star is not
expected to show clear oscillations either (and is most likely a dwarf or
too faint giant with a white noise dominated spectrum).
Uncertainties were computed with a weighted centroids method that depends
on the frequency resolution in the PS2.  Because we used the same number of
orders around \numax\ for most stars, the relative uncertainties are 
generally very similar for the stars.  A2Z returned \numax\ and
\dnu\ results for 673 stars. 

{\bf SYD}: This pipeline is described in detail in \citet{Huber09}, and
uses Monte-Carlo simulations to generate statistically perturbed spectra
from which measurement uncertainties are estimated as described in \citet{Huber11a}.
The pipeline does not include a statistical assessment of whether
oscillations are present in the data. Hence, to automate the selection of
robust detections we impose a number of post processing criteria.  
First, stars are kept only if the power spectrum is steep
enough to show the presence of granulation, using the same criteria as A2Z.
Secondly, the correlation between an empirical average model spectrum and
the slightly smoothed collapsed power spectrum of the central four orders
need to be above a certain cut-off value (to ensure a clear mode pattern
revealing \dnu). The correlation process follow that of
\citet{Stello16b}, and like CAN, the cut-off is informed by visual
inspection of stars that are accepted versus rejected.
Thirdly, the stars need to have a \dnu\ within 30\% of the average expected
\dnu\ from the \dnu-\numax\ relation of \citep{Stello09a}. This cut was
chosen based on the natural spread seen among the high signal-to-noise
\kepler\ red giant sample \citep{Stello13}.
Finally, stars need to have an oscillation amplitude within a factor of two of
the average fiducial amplitude-\numax\ trend seen for spectroscopically 
confirmed red giants from \citet{Stello15}.
SYD returned results for all stars thought to provide robust values for
both \dnu\ and \numax\ (559 stars). 

{\bf BAM}: Like CAN, BAM uses a Bayesian MCMC scheme for fitting the
background and the oscillation excess power hump in the frequency spectrum,
and for estimation of the measurement uncertainties (Zinn et al. in
prep.).
It first fits a two-component noise background (plus white noise) whose
resulting parameters are subsequently used as priors on a fit to the
oscillation excess, modeled as a Gaussian with amplitude, width, and center
(\numax) as free parameters. One set of priors are calculated from
empirical correlations  between the background parameters and \numax;
another prior is placed on the timescale ratio between the two background
components; and a final set of priors link \numax\ to the Gaussian
amplitude and width.  These priors are then used to derive the (total
prior) probability of finding the measured value of each fitted parameter. 
After fitting the model (of noise plus excess) to the power spectrum, BAM
selects a star as an oscillator if: 1) the signal-to-background ratio
at \numax\ is greater than 3.0, and 2) the total prior probability is
within the 4-sigma confidence interval. 
Subsequently, \dnu\ is determined by a least-squares fit to the
folded spectrum using the 4\dnu-wide \numax-centered section of the
spectrum with \dnu\ as the folding frequency. The folded radial,
dipole, and quadrupole mode regions are each modeled with a Lorentzian function.  
The uncertainty in \dnu\ is estimated as the quadrature sum of the
uncertainties on the mode widths and locations. \dnu\ is returned for stars
only if the probability of \dnu\ and the mode visibilities are within
$4\sigma$ of their expected values based on priors. BAM returned
\numax\ for 951 stars and \dnu\ for 759 stars. 

Some of the pipelines used here have previously been compared on the basis of
\kepler\ data.  Although they have improved since, and changes has been made to
accommodate K2-specific issues, we refer the reader to such comparison
papers for additional reading \citep{Hekker11a,Hekker12}.

\section{Seismic results}\label{results}
In Table~\ref{tab1} we list the total number of stars for which each pipeline
detected the presence of oscillations, meaning at least a measurement of
\numax. When combined, we found 1262 unique stars with at least one 
pipeline reporting detection.  In Table~\ref{tab1b} we list the total
number of stars for which the pipelines also detected \dnu.
\begin{table}
{\footnotesize
\begin{center}
\caption{Number of red giants with detected \numax\ and comparison
  with visual inspection. \label{tab1}}
\begin{tabular}{lrrrrrrr}
\tableline\tableline
              Pipeline &   CAN  &   COR  &  BHM  &   A2Z  &   SYD  &  BAM     &  Comb. \\
\tableline
               Total   &  1106  &   778  &  1031  &   673  &   559  &   951    &  1262  \\
\tableline
               Yes     &  1017  &   745  &   980  &   638  &   537  &   905    &  1100  \\
\multicolumn{1}{r}{\%} & (92.0) & (95.8) & (95.1) & (94.8) & (96.1) & (95.2)   & (87.2) \\
               Maybe   &    67  &    28  &    36  &    26  &    17  &    34    &   110  \\  
\multicolumn{1}{r}{\%} &  (6.1) &  (3.6) &  (3.5) &  (3.9) &  (3.0) &  (3.6)   &  (8.7) \\
               No      &    22  &     5  &    15  &     9  &     5  &    12    &    52  \\
\multicolumn{1}{r}{\%} &  (2.0) &  (0.6) &  (1.5) &  (1.3) &  (0.9) &  (1.3)   &  (4.1) \\
\tableline
\end{tabular}
\tablenotetext{}{The `Total' numbers are split into the three visual
  classification categories shown both as absolute numbers and in percent
  for each pipeline. Due to rounding, the percentages do not necessarily
  add to exactly 100\%. `Comb.' refers to stars with at least one pipeline
  detection.} 
\end{center}}
\end{table}

\begin{table}
{\footnotesize
\begin{center}
\caption{Number of red giants with detected \dnu. \label{tab1b}}
\begin{tabular}{lrrrrrrr}
\tableline\tableline
              Pipeline &   CAN  &   COR  &  BHM  &   A2Z  &   SYD  &  BAM     &  Comb. \\
\tableline
               Total   &   583  &   778  &   198  &   673  &   559  &   759    &  1184  \\
\tableline
\end{tabular}
\tablenotetext{}{`Comb.' refers to stars with at least one pipeline
  detection.}
\end{center}}
\end{table}

The \dnu\ and \numax\ values from each pipeline are listed in
Table~\ref{tab2}.
All stars with no detection of oscillations, are listed in
Table~\ref{tab3} by their EPIC ID,
$\mathit{Kp}$, and $\log g$-range classification probability from stellar
population modeling using \galaxia. They are a source of mostly dwarfs with a well
characterised selection function (Section~\ref{targetselect}).

\begin{sidewaystable}
{\footnotesize
\begin{center}
\caption{Seismic results and visual detection verification of 1262 giants detected
  by at least one pipeline. `Y'=Clear detection; `M'=Maybe detection;
  `N'=No detection; `O'=Clear detection but \numax\ off. All frequencies
  and their uncertainties are listed in \muhz. \label{tab2}} 
\begin{tabular}{lcrrrrrrrrrrrrrrrrrrrrrrrr}
\tableline\tableline
EPIC ID  & Detec. &                  \multicolumn{4}{c}{CAN}                &                  \multicolumn{4}{c}{COR}                &                  \multicolumn{4}{c}{BHM}               &                  \multicolumn{4}{c}{A2Z}                &                  \multicolumn{4}{c}{SYD}                &                  \multicolumn{4}{c}{BAM}               \\
         &        &  \multicolumn{2}{c}{\dnu} & \multicolumn{2}{c}{\numax}  &  \multicolumn{2}{c}{\dnu} & \multicolumn{2}{c}{\numax}  &  \multicolumn{2}{c}{\dnu} & \multicolumn{2}{c}{\numax}  &  \multicolumn{2}{c}{\dnu} & \multicolumn{2}{c}{\numax}  &  \multicolumn{2}{c}{\dnu} & \multicolumn{2}{c}{\numax}  &  \multicolumn{2}{c}{\dnu} & \multicolumn{2}{c}{\numax} \\
\tableline
201121245&    Y   &   15.977      &   0.029   &   198.113     &   2.838     &      14.887  &  0.174     &  202.510  &  4.180          &   16.100  &  0.269        &  187.286  &  3.391          &      0.000   & 0.000      &   0.000  &  0.000           &    16.046 &    0.056      &   199.530  &  2.534         &      0.000  &  0.000      &    0.000  &  0.000     \\
201126368&    M   &    0.000      &   0.000   &     0.000     &   0.000     &       4.211  &  0.075     &   47.470  &  1.240          &    0.000  &  0.000        &    0.000  &  0.000          &      0.000   & 0.000      &   0.000  &  0.000           &     0.000 &    0.000      &     0.000  &  0.000         &      4.579  &  0.046      &   42.966  &  0.966     \\
201126489&    Y   &   17.174      &   0.034   &   226.041     &   3.462     &       0.000  &  0.000     &    0.000  &  0.000          &   16.851  &  0.879        &  223.491  &  5.228          &     17.200   & 0.108      & 226.274  & 15.172           &    17.160 &    0.035      &   224.334  &  3.154         &     17.003  &  0.026      &  224.423  &  3.129     \\
...      &   ...  &  ...          & ...       &  ...          & ...         &  ...         & ...        &  ...      & ...             &  ...      & ...           &  ...      & ...             &  ...         & ...        &  ...     & ...              &  ...      & ...           &  ...       & ...            &  ...        & ...         &  ...      & ...         \\   
\tableline   
\end{tabular}
\tablenotetext{}{full table is available electronically}
\tablenotetext{}{Stars that do not follow the K2 GAP target selection
  (Sect~\ref{targetselect}) are listed with their EPIC IDs made negative.}
\end{center}}
\end{sidewaystable}

\begin{table}
{\footnotesize
\begin{center}
\caption{List of stars without seismic detection (almost all dwarfs).
  Probabilities are divided into four ranges of $\log g$: $\log g <1.9$,
  $1.9<\log g<3.2$, $3.2<\log g<4.2$, $\log g>4.2$ (see Sect.~\ref{results}).
  \label{tab3}} 
\begin{tabular}{lcrrrrrrrrrrr}
\tableline\tableline
EPIC ID  &  $\mathit{Kp}$     & \multicolumn{4}{c}{Probability [\%]} \\
         &  [mag]    & $P_{<1.9}$ & $P_{1.9-3.2}$ & $P_{3.2-4.2}$ & $P_{>4.2}$  \\
\tableline
 201122454&  13.85   & 4   & 1   & 0   &*95\\
 201123619&  15.21   & 2   & 0   & 0   &98\\
 201124136&  12.24   &16   &19   & 2   &63\\
...       &  ...     & ... & ... & ... & ...\\

\tableline
\end{tabular}
\tablenotetext{*}{Due to rounding errors, the printed probabilities do not
  necessarily add to exactly 100\%.}
\tablenotetext{}{(full table is available electronically)}
\end{center}}
\end{table}

In addition to the following verification and completeness analyses, we show
a series of figures in the Appendix comparing the results from the 
individual pipelines to aid the use of this rich and versatile data set.
The figures highlight statistical properties of the results including the
natural biases between pipelines arising from the fact that each pipeline
measures and hence defines the seismic observables in slightly different
ways.  Users for which the magnitude of these biases are important for
their science are strongly encouraged to familiarise themselves with the
specific methods adopted by the pipelines and their comparisons (see
references in Sect.~\ref{pipelines}).

\subsection{Visual verification}
We visually inspected power spectra 
for all 1262 pipeline-detected giants, and assigned them to one of three
categories `Yes', `Maybe', or `No', according to whether they showed
oscillations and/or granulation. 
This is obviously a subjective approach, which in a sense takes a continuum
of probabilities of a star showing oscillations and puts that into three
bins. However, we did it such that `Yes' meant we felt confident the star
oscillated, while `No' meant we were confident it did not oscillate below
the Nyquist frequency.  
\begin{figure*}
\includegraphics[width=17.6cm]{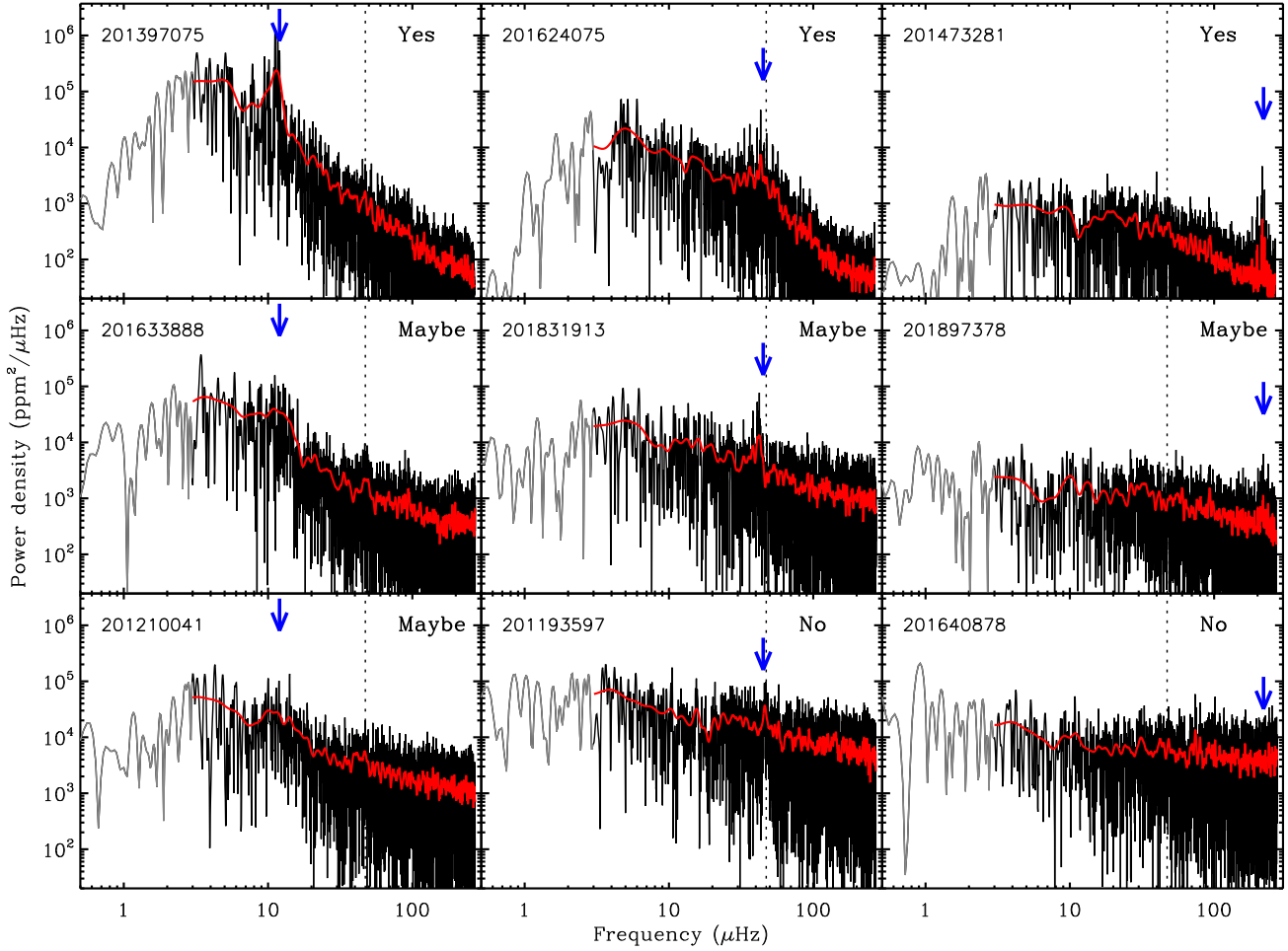}
\caption{Example power spectra for three representative \numax\ values
  (left panels: \numax\ $\sim 10\,$\muhz; centre panels: \numax\ $\sim
  45\,$\muhz; right panels: \numax\ $\sim 220\,$\muhz), and three different
  visual verification categories (top panels=Yes; middle panels=Maybe;
  bottom panels=No). There was no 'No'-category stars with a low
  \numax\ measurement by the pipelines, hence the bottom left panel shows a
  Maybe case.  The raw spectra are shown in black and the part affected 
  by the high-pass filtering in grey. Heavily smoothed spectra are shown in
  red.  Approximate \numax\ values detected by the pipelines is
  indicated by the blue arrows, and the dotted vertical lines shows the
  spacecraft thruster firing frequency. Each star is indicated by their
  EPIC ID \citep{Huber16}.
\label{example_spectra}} 
\end{figure*} 
Figure~\ref{example_spectra} shows a sample representative of low
(left panels), medium (centre panels), and high (right panels)
\numax\ detections (by the pipelines). For each \numax\ case we show the
resulting three different visual detection categories; except for the low
\numax\ case, where there were no stars categorised as `No' detections, for
which we instead show a 'Maybe' case that straddle the No category.

To test the subjectivity of our visual inspection, we had two people look
through the power spectra independently.  Cross checking their results showed not a
single case where one person said 'No' and the other said 'Yes', and there
were only five borderline cases between neighboring categories, where one
said 'Yes' or 'No' and the other said 'Maybe'.  
We therefore regard the result of the visual inspection as highly robust. 
A similar exercise had been implemented by \citet{Stello13}, also showing
consistency between three independent people.   
We include the result of the visual detection verification in
Table~\ref{tab2} 
and summarise how each pipeline performs relative to our visual inspection
in Table~\ref{tab1} (in pure counts and percent). 
It is worth noting that despite the larger number of stars 
returned by CAN, BHM, and BAM, those samples seem to
be of similar quality as the rest (high `Yes' and low `No' fractions). 

\begin{figure*}
\includegraphics[width=17.6cm]{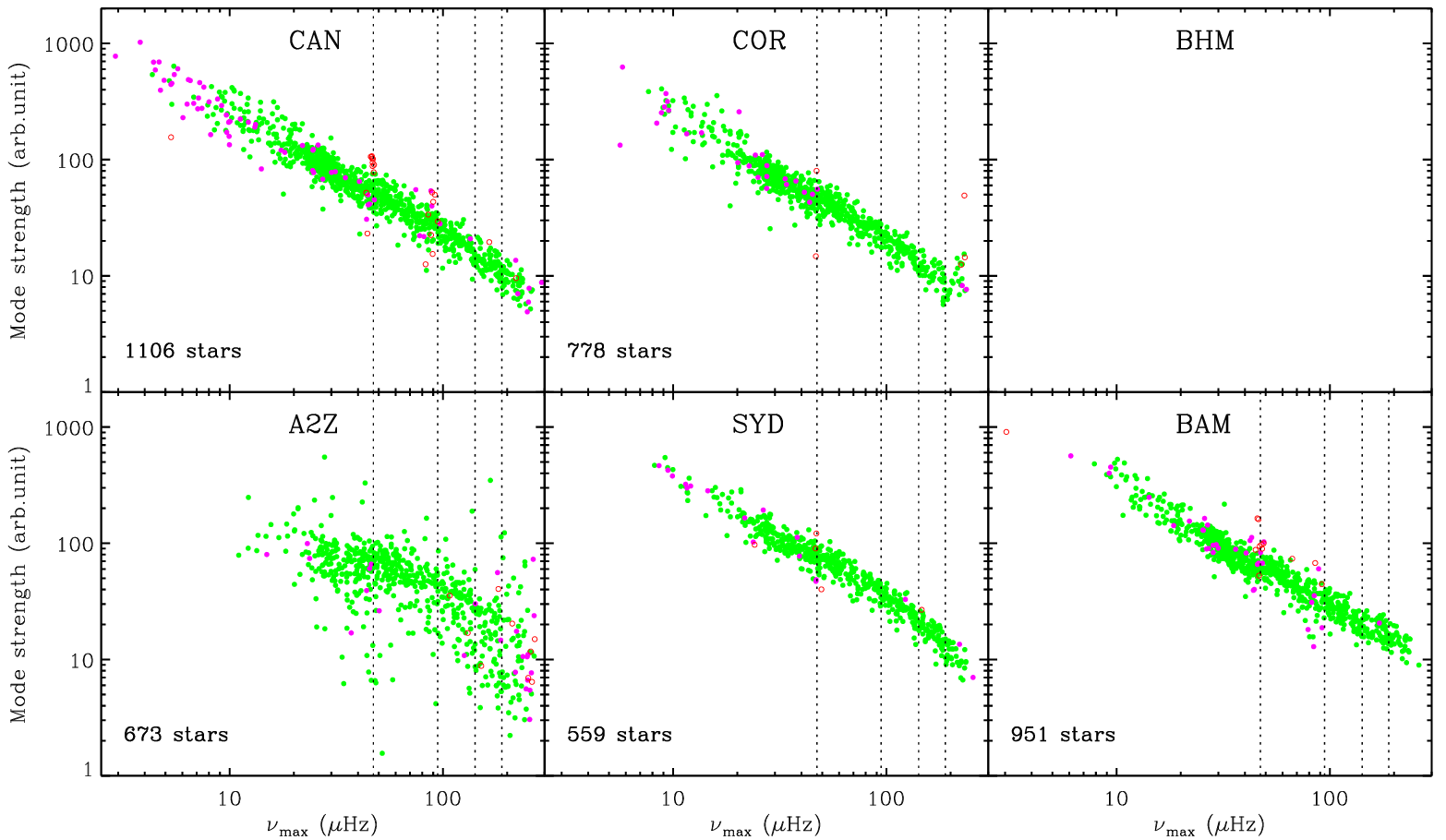}
\caption{Mode strength versus \numax\ from each pipeline. Mode strength
  is synonym for either oscillation amplitude or some other measure of the
  oscillation excess power (BHM did not return such values).  
  The K2 thruster firing
  frequency and its first few harmonics are shown by dotted lines.
  Green filled: 'Yes' detections,
  Magenta filled: 'Maybe' detections, and 
  Red open: 'No' detections by visual inspection.
\label{amp-numax}} 
\end{figure*} 

Of the 1262 giants detected by at least one pipeline, we found 1100 in the
`Yes' category, 110 were assigned as `Maybe', and 52 `No' detections.  
If we assume the `Maybe' cases are all genuinely oscillating giants, just
with low S/N ratios or with \numax\ outside the 10-270\muhz\ range,
this would suggest we have up to $\sim 1210$ detections.   

To further double check that the confirmed detections were in line in
terms of their \numax\ values, we compared pipeline values with visual
values. The latter was found by manually placing (and mouse-clicking) a
cursor on interactive plots of power spectra at the location of \numax. 
The intrinsic uncertainty in performing this manual \numax\ localisation,
of up to 30\%, meant we could safely use them to identify clear outliers.  It
was encouraging to see only a handful of pipeline results were indicated as
outliers.  

\subsection{Ensemble verification}\label{ensembleverification}
One way to check that the \dnu\ and \numax\ measurements originate from
oscillations and not from spurious humps of excess power is to verify that
the amplitude, or some sort of similar measure of mode oscillation
strength, show a correlation with \numax\ as expected from scaling
relations \citep[e.g.][]{Stello07}, and empirically demonstrated for large
samples of giants \citep[e.g.][]{Mosser10,Huber10}.
In Figure~\ref{amp-numax} we show mode strength versus \numax\ for the
pipelines that return both quantities.  
All panels show correlations indicative of detections due to
oscillations.   
We also indicate the visual verification results by different colored symbols.
It is reassuring to see that the `No' detections (red open symbols)
fall predominantly in the narrow frequency ranges around the spacecraft's thruster
firing frequency and its harmonics or very close to the Nyquist frequency.
Due to an improvement in the spacecraft altitude control since campaign 3, we
expect the thruster firing will have negligible impact on the data for 
later campaigns as evidenced by the clean C5-based data in \citet{Stello16M67}.  
The `Maybe' category is spread out more (magenta filled symbols), but often
at extreme frequencies 
pushing the limit of what can be measured with the length and sampling of the
K2 data.  Or they are near the artifacts from thruster firings; although a
significant fraction are between 20 and 30\muhz\,
which are probably helium-core burning stars or possibly artifacts at half
the thruster 
firing frequency.   These results show that the visual inspection provides
valuable complementary information, which suggests that the pipelines
only struggle in a limited range and mostly extreme ends of the parameter space.

\begin{figure*}
\includegraphics[width=17.6cm]{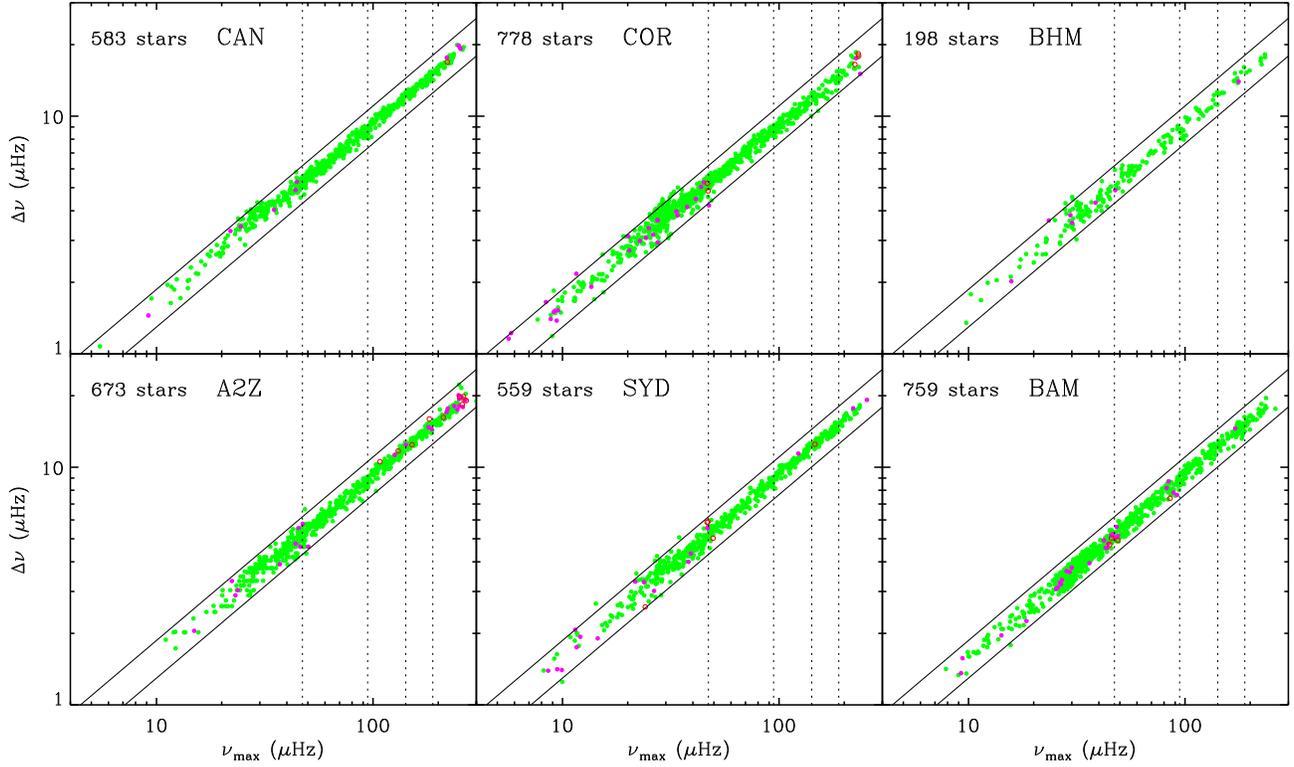}
\caption{\dnu-\numax\ relation from each pipeline.  The K2 thruster firing
  frequency and its harmonics are shown by dotted lines.
  To guide the eye, black lines show $+/-$ 20\% brackets from the
  \dnu-\numax\ relation by \citet{Stello09a} outside which one would
  normally not find robust detections of high S/N (e.g. from 4-yr
  \kepler\ data). 
  Green filled: 'Yes' detections,
  Magenta filled: 'Maybe' detections, and 
  Red open: 'No' detections by visual inspection.
\label{dnu-numax}} 
\end{figure*}

To further investigate whether the seismic measurements agree with
expectations, we show the \dnu-\numax\ relation in Figure~\ref{dnu-numax},
which is expected to be relatively tight both from scaling relations
\citep{Stello09a} and as demonstrated for large samples of giants
\citep[e.g.][]{Hekker09,Huber10}.  
From \kepler\ data we generally do not see stars outside the $\pm20\,$\%
range around the mean \dnu-\numax\ relation (solid lines), which roughly encompasses a
mass spread of 0.5-2.5\msol.  


\section{Completeness of seismic sample}\label{detectbias}
In the following we investigate to what extent our resulting sample of
oscillating giants is affected by detection bias. 
Unlike \citet{Stello15} we do not have prior spectroscopic knowledge of the
evolutionary state of our stars 
to guide this analysis.  For a given non-detection we can
therefore not be certain if it is outside the detectable frequency range or if
it is just a very low S/N case.  However, we have a much larger set of
stars than in \citet{Stello15}, which should enable us to make inference on
the presence of likely detection biases by comparing our samples of \dnu\
and \numax\ detections, 
and in turn compare the latter sample with
a synthesized population. 
The obvious caveat with the synthesized population will be its
model dependency.  But as long as the model is not grossly inaccurate we can
reasonably assume only small detection biases could remain unnoticed. 

\subsection{Completeness of \dnu\ sample}
\begin{figure*}
\includegraphics[width=17.6cm]{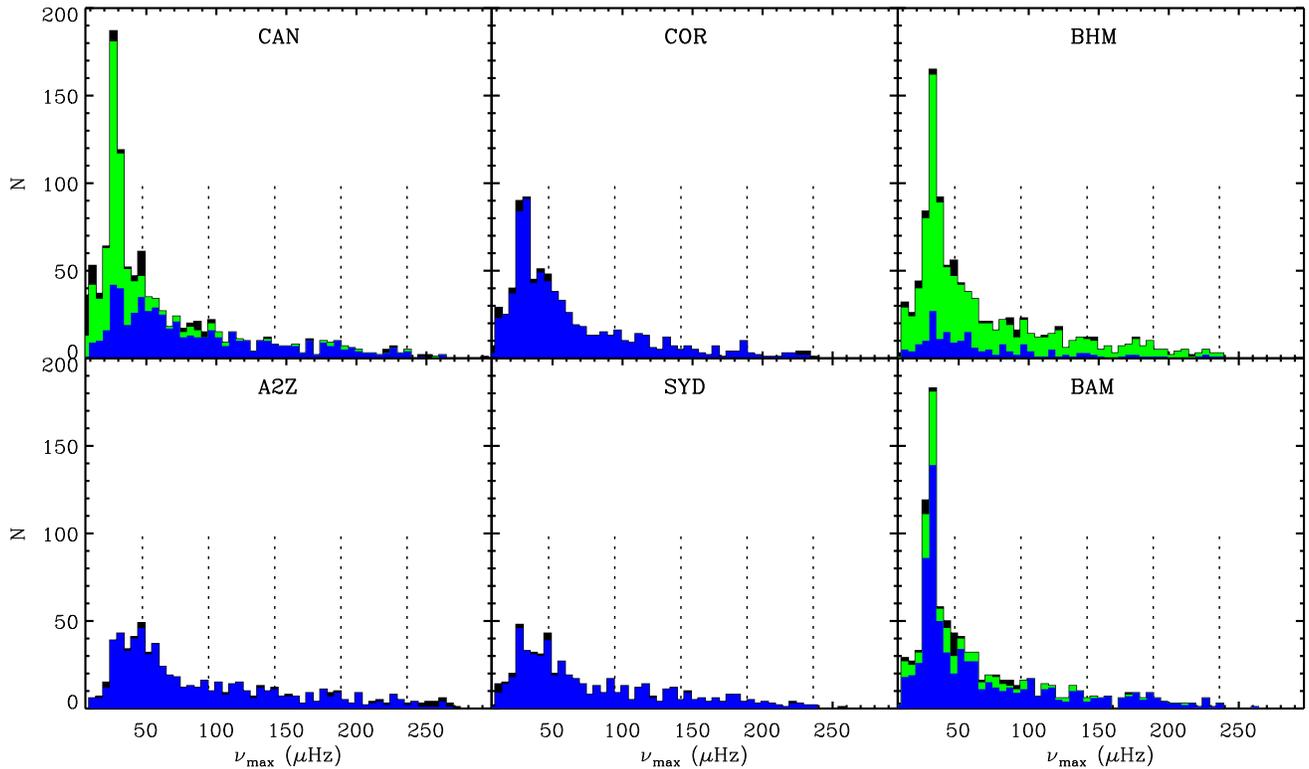}
\caption{Distribution of \numax\ for detected red giants. Black shows
  all stars with a measured \numax\ value, while green shows only
  the visually verified `Yes' detections. Blue represent the
  verified detections for which \dnu\ is also measured (equal to green
  for COR, A2Z, and SYD). 
\label{numax-histo}} 
\end{figure*}

Comparing the number of stars between Figure~\ref{amp-numax} and
~\ref{dnu-numax} plotted in each panel, illustrates that more stars show evidence of
oscillation excess power (a \numax\ measurement) than a clear frequency
pattern from which \dnu\ can be determined (see also Tables~\ref{tab1} and
\ref{tab1b}).  This is an inevitable 
consequence of the length of the time series, which is not long enough to
fully resolve the mode patterns in some of the stars.  In
Figure~\ref{numax-histo} we show the histogram of stars with a visually verified \numax\
measurement (green), compared to those where we also have a
\dnu\ measurement (blue).  It
is evident that the problem of identifying \dnu,
affects the helium-core burning stars the most, particularly stars with
\numax\ around 30\muhz. This arises because they show 
more complex frequency spectra, and mode power is spread into more
frequency bins,
which are therefore each of lower amplitude, resulting in less clear
mode patterns.
Based on Figure~\ref{numax-histo} we conclude the sample of stars with both
\dnu\ and \numax\ measurements is biased at least for helium-core burning
stars.  It is not clear if that applies only to the helium-core burning
stars that have
\numax\ $\sim 30\,$\muhz.  The \dnu\ detection bias 
is more severe for some pipelines where we see a clear lack of stars in
the blue curve compared to the green curve (e.g. CAN), potentially
affecting all stars with \numax\ $\lesssim 50\,$\muhz. 
Looking at \citet{Hekker11} (their Figure 2) it
seems that for 50-100 day \kepler\ time series, which brackets the
80-day K2 C1 series, there is a deficit of stars
with \dnu\ in the range 3-5\muhz\ (equivalent to \numax\ $\sim 30\,$\muhz)
relative to the expected smooth decrease 
in detections for decreasing \dnu\ (or \numax).  
This is in qualitative agreement with what we see here for K2.
We therefore recommend that the \dnu\ detection bias for the potentially
affected stars (\numax\ $\lesssim 50\,$\muhz) be quantified specifically 
for K2 using the same pipelines as presented here before the 
\dnu\ results of these stars are used for population
studies. However, the reported \dnu\ could of course still be
used safely for other purposes (or indeed for population studies
restricted to stars in the unaffected \numax\ range
(Figure~\ref{numax-histo}). 

\subsection{Completeness of \numax\ sample}
\subsubsection{Individual pipeline results}
Here, we will compare our number of detected giants with what one would
expect from synthesizing our target sample.  
We note that while the total number of detected giants is up to 1210, they
were found among all 8,630 observed stars in the K2 GAP list, rather than 
only the 8,385 that follow the K2 GAP selection
(Section~\ref{targetselect}). Hence, about 30 of the  
detected giants are serendipitous from other K2 proposals, which should be
subtracted before comparing with a synthesized sample\footnote{In
  Table~\ref{tab2} we have marked the stars that do not following the K2
  GAP selection}. 
We use the Galaxy synthesis tool, \galaxia\ to synthesize our target sample
of 8,385 stars by imposing the selection described in
Section~\ref{targetselect}, and the \numax\ detection range of 10-270\muhz.
This results in an initial $\sim 1650$ expected giants (within Poisson
error) in the detectable \numax\ range, which is significantly larger than
the number of giants actually detected.  
Hence, we want to investigate whether there might be any
significant biases with respect to signal and/or noise.  The signal is the
oscillation amplitude for which \numax\ is a good proxy (Fig.~\ref{amp-numax}),
while the noise can be approximated by brightness, $\mathit{Kp}$.
\begin{figure*}
\includegraphics[width=17.6cm]{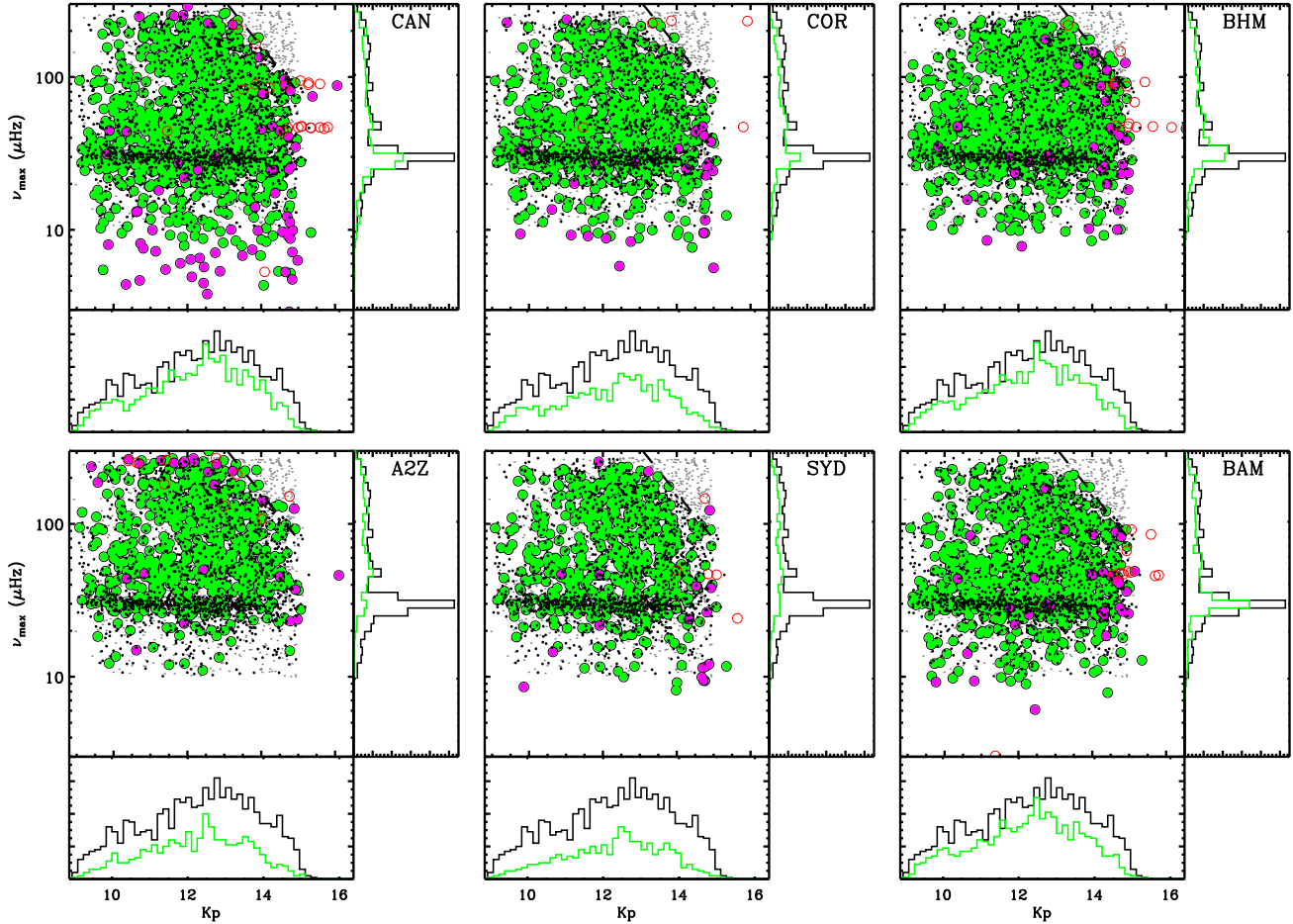}
\caption{\numax versus $\mathit{Kp}$ of seismic sample from each pipeline.
  Green filled symbols shows only stars verified visually as clear detections,
  while magenta filled symbols are maybe detections, and red open symbols
  are non-detections. 
  Small grey dots show the \galaxia-synthesized population. Black dots show
  only those \galaxia\ stars with more than 95\% probability of detection
  according to the \citet{Chaplin11} formulation. 
  The dashed line shows a fiducial detection limit (the same line is used
  in all panels for comparison), such that oscillations in stars above 
  the line can not be detected 
  (\numax$_\mathrm{,detect} < 2.6\times 10^6 \cdot 2^{-\mathit{Kp}}\,$\muhz).
  The distributions of the green symbols and black dots are shown as
  histograms in \numax\ and $\mathit{Kp}$.
\label{numax-kpmag}} 
\end{figure*} 
In Figure~\ref{numax-kpmag} we therefore show \numax\ and $\mathit{Kp}$ for all visually
inspected detections (green, magenta, and red symbols).  The grey dots in
the background show the synthesis population of 1650 expected giants.
It is clear from this comparison that we can not detect oscillations in stars
in the upper right corner, which are the 
faintest (more noisy), and the least evolved stars (highest \numax\ and
hence lowest intrinsic oscillation amplitudes).   

To assess if this apparent `lack' of faint low luminosity red giants is
expected, we need to take the oscillation detectability into account.  
We derive the detectability 
following 
the formulation by \citet{Chaplin11}.  The stellar properties are given as
part of the output from \galaxia, from which the 
oscillation and granulation signal is predicted.  Different to
\citet{Chaplin11}, we use the more up-to-date relation for oscillation
amplitude by \citet{Stello11b}, which is empirically calibrated by open
cluster red giants.  
We also need to re-calibrate the noise estimates
compared to the \kepler\ results \citep{Jenkins10} adopted by
\citet{Chaplin11} to incorporate the K2-specific noise properties. 
\begin{figure}
\includegraphics[width=8.0cm]{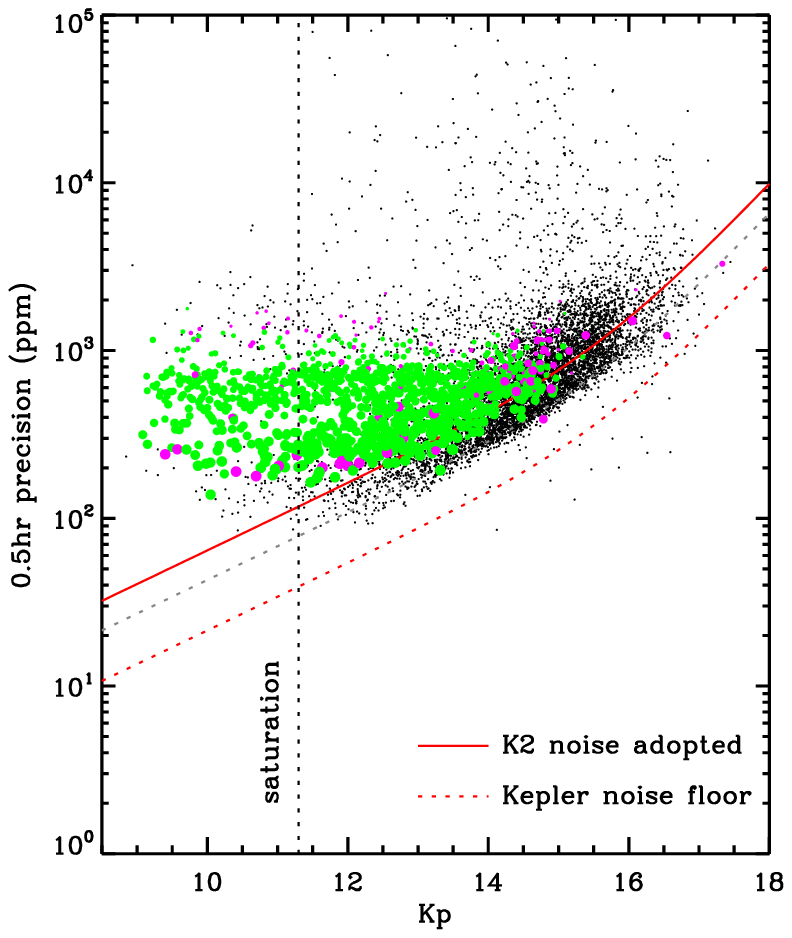}
\caption{Time series scatter (standard deviation) versus $\mathit{Kp}$ of all
  K2 GAP targets (black dots).   
  Green filled  symbols shows stars verified visually as clear detections,
  while magenta filled symbols are maybe detections. 
  Symbol size scales with \numax.  
  Vertical dotted line indicates the saturation limit. The dotted red curve
  shows the \kepler\ noise floor from \citet{Jenkins10}. The grey dotted
  line is shifted up by a factor of two.  The red solid curve
  illustrates our adopted K2 noise, obtained by raising the \kepler\
  noise floor by a factor of three. 
\label{rms-kp}} 
\end{figure} 
To illustrate the K2-\kepler\ noise difference, we show in
Figure~\ref{rms-kp} the measured scatter in the time series for all stars in our
sample (black dots).  For comparison we show the noise floor of
the equivalent \kepler\ time series scatter from \citet{Jenkins10} (dotted
red curve), which matches the K2 noise floor quite well (grey dotted
curve) when shifted up be a factor of two.  However, this noise floor is
clearly a lower limit even for photon noise dominated stars
($\mathit{Kp}\gtrsim14$), which for the vast majority show a scatter that
ranges a factor of four. We therefore adopt a larger and more representative
noise level for K2 of three times the \kepler\ level (solid red curve).

Now, taking our derived detectability into account, we expect to detect oscillations
in only $\sim 1420$ of the synthesis sample (Figure~\ref{numax-kpmag}, black
dots).  This synthesis sub sample shows the same lack of stars as our
observations in the top right corner in the Figure~\ref{numax-kpmag} main panels. 
For added clarity we also show histograms of the observed and synthesis
distributions as function of \numax\ and $\mathit{Kp}$.  The $\mathit{Kp}$ distributions
demonstrate a drop in stars occurs beyond $\mathit{Kp}=12.5$-12.7 (green curve),
which is similar to what is expected from the \galaxia\ distribution
peaking at $\mathit{Kp}=12.7-13.0$ (black curve).  
There seem to be a tendency for all pipelines to claim detections where no
detection was verified by the visual inspection for stars fainter than
$\mathit{Kp}\sim 15$ (red open symbols).  We therefore caution using these
seismic results of 
stars fainter than $\mathit{Kp} \simeq 14.5$ for population studies due to potential
completeness issues in the current sample of results. However, we note that the lack of luminous faint ($\mathit{Kp}>15$)
giants in our C1 sample prevents us from reaching definitive conclusions
about detection bias towards the faint end. 

\subsubsection{Combined pipeline results}
\begin{figure*}
\includegraphics[width=17.6cm]{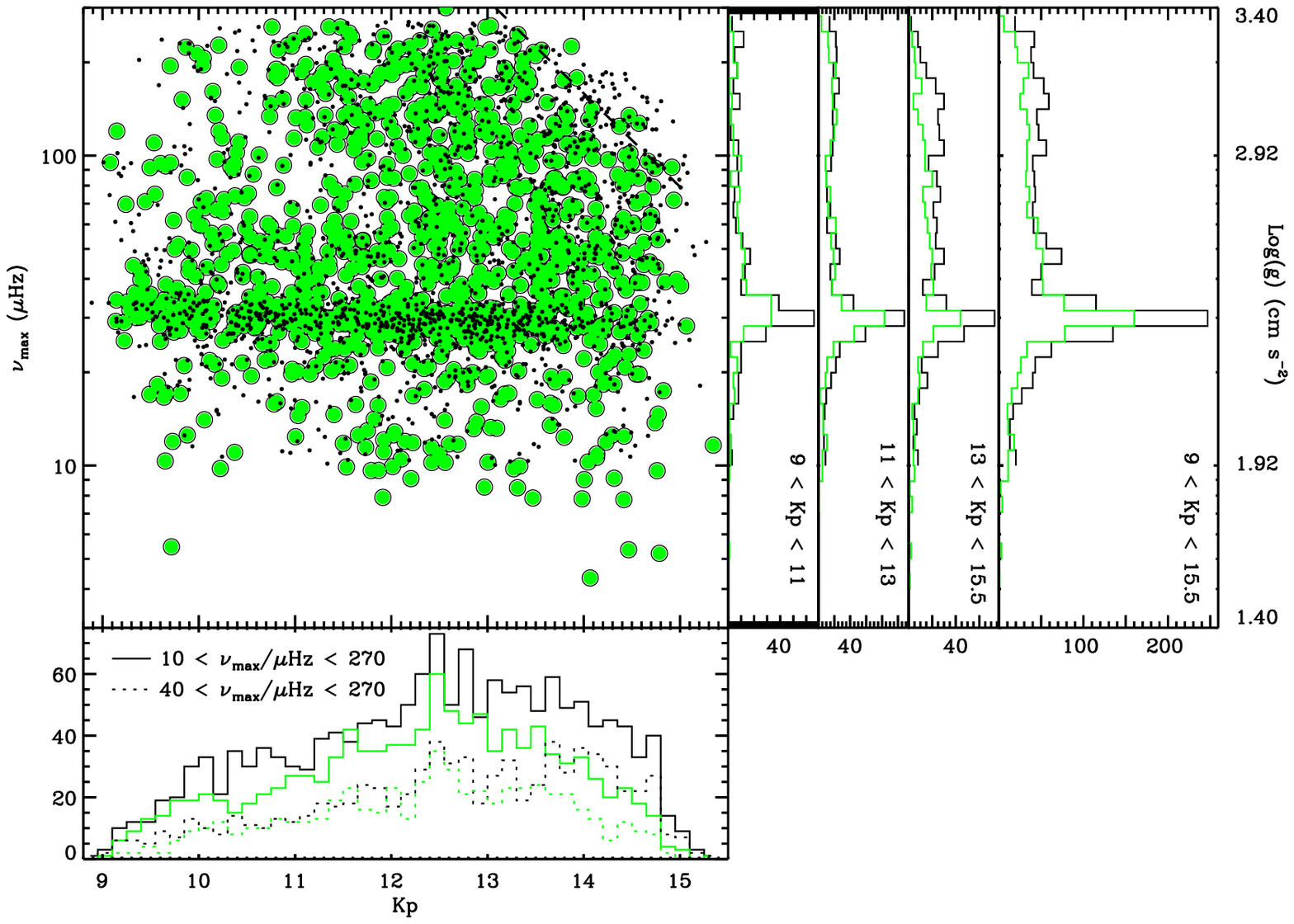}
\caption{\numax versus $\mathit{Kp}$ of combined seismic sample from all pipelines.
  The axis on the far right shows the approximate $\log\,g$ range.
  Green filled symbols/lines show only stars verified visually as clear detections.
  Black symbols/lines show
  the \galaxia-synthesized population that have more than 95\% probability
  of detection according to the \citet{Chaplin11} formulation. 
  The dashed line in the main panel shows a fiducial detection limit
  (\numax$_\mathrm{,detect} < 2.6\times 10^6 \cdot 2^{-\mathit{Kp}}\,$\muhz),
  beyond which the signal-to-noise ratio is too low.
  The distributions of the green symbols and black dots are shown as
  histograms in \numax\ for different magnitude ranges and in
  $\mathit{Kp}$ for different \numax\ ranges.
\label{numax-kpmag-comb}} 
\end{figure*}

Now we turn to the combined set of results, and show the sample
distribution in \numax-$\mathit{Kp}$ space in Figure~\ref{numax-kpmag-comb}.  
Here we acknowledge that
different pipelines have different strengths and by compiling all the
stars for which at least one pipeline detected oscillations, 
we are more likely to obtain as many stars as possible in our sample.
Hence, our sample will be more likely to contain, as closely as possible to, all
detectable oscillating stars. 
For this part of our investigation we adopted \numax\ values first sourced
from CAN, then we added those from A2Z not already in the CAN sample, then
BHM, SYD, COR, and finally BAM, following the order of the pipeline with
the most to least unique detections.  
By construction, not all stars in the compiled list
have a \numax\ value from the same pipeline. However, for the purpose of
investigating the completeness of the population as a whole, the relatively
small \numax\ biases between different pipelines plays no important role
(Appendix, Figures~\ref{appfig9},~\ref{appfig10}).   

To use the sample for population studies one needs to take into account
that we can not detect stars above the fiducial dotted curve, expressed by  
\numax$_\mathrm{,detect} < 2.6\times 10^6 \cdot 2^{-\mathit{Kp}}\,$\muhz.
We note that the curve seems to have a slightly steeper dependence on $\mathit{Kp}$
than predicted (see black dots).  The predicted dependence is close to
$\propto 2^{-0.4\mathit{Kp}}$ as expected from the photon noise limit, and
we attribute the steeper observed dependence to the fact that our adopted
noise estimate becomes increasingly underestimated towards fainter stars;
as opposed to bright stars where the photon noise plays no significant role
in the detectability.  In order words, our simple noise prescription does
not take full account of the gradual transition between the two noise regimes
and the intrinsic spread in the noise from star to star.  It is therefore
also difficult to make an accurate model prediction of the absolute number of
expected detections.  Changing the adopted noise level by about 30\%,
changes the number of expected detections by about 100 (with larger noise
resulting in fewer expected detections).  

From the distributions shown as histograms of different parts of the
\numax-$\mathit{Kp}$ parameter space we see that the discrepancy
between observed and predicted number of stars seems to be more pronounced
for \numax\ $\lesssim 40\,$\muhz, which affects many helium-core burning
stars. Here it shows about a 30\% difference
(Figure~\ref{numax-kpmag-comb}, right panels).  This difference could be
due to both observation bias and errors in the Galactic model prediction
including predictions of oscillation amplitude and noise.  While
observational detection bias of \numax\ would not be unexpected for the
faintest stars, it is surprising to see the quite pronounced discrepancy
for the bright stars.  
There were no strong indications of such pronounced discrepancies for
the \kepler\ red giant sample based on two years of data \citep{Sharma16}.
However, due to the unknown selection function of the \kepler\
giants direct comparisons can not be made with what we see here for the 
K2 C1 sample.  
In the bottom histogram we also do not see any strong dependence on
$\mathit{Kp}$ for the  
difference between observed and predicted star counts, except perhaps a
small tendency of relatively fewer observed stars fainter than
$\mathit{Kp}\sim 13.5-14.0$, mostly arising from the before-mentioned
slight difference in observed and predicted bias around the sloping dashed
line (faint low luminosity stars).
Also in this histogram, we see that the star counts agree quite well 
for \numax\ $> 40\,$\muhz\ (dotted curves) in agreement with the above
discussion.  


In summary, we see an expected (and largely predicted) detection bias for
faint low luminosity red giant branch stars and an apparent, but still not
fully understood, difference between observed and predicted star counts for
red clump stars across all magnitudes.

\subsection{Comparison to EPIC classifications}
Another source to investigate completeness are the stellar classifications
provided in the Ecliptic Plane Input Catalog \citep{Huber16}. While the
EPIC uses \galaxia\ models for stellar parameter inference, the
classifications are based on observational data such as reduced proper
motions, colors, and spectroscopy, hence providing a more empirical test of
the completeness of the seismic sample. Based on simulations and
comparisons to the Kepler Stellar Properties Catalog
\citep{Huber14,Mathur16}, the dwarf/giant misclassification fraction in the
EPIC is expected to be $\sim$\,5\%. 

Of the 1262 stars with a \numax\ value reported by at least one pipeline,
all but one are classified in the EPIC. Of these, $\sim$\,9\% are
classified as dwarfs in the EPIC ($\log g > 4$) and $\sim$\,2\% are
classified as cool dwarfs ($\log g > 4$, \teff\ $<4000\,$K). We expect that
the hotter dwarfs (\teff\ $>4000\,$K) are significantly affected by
classification errors in the EPIC, and hence account for most of the 5\%
misclassification fraction mentioned above. Inspection of a 2MASS
color-color diagram showed that at least half of the dwarfs with
\teff\ $<4000\,$K indeed have colors consistent with cool dwarfs. We
therefore conclude that the fraction of seismic stars in our sample which
are in fact dwarfs (with the seismic signal introduced e.g. by blended
giants) is $\sim$\,2\%.    

Using the EPIC to estimate the number of missing seismic detections is more
difficult, since the typical EPIC $\log g$ uncertainties on the RGB can be
as high as 0.3\,dex. Nevertheless, the fraction of C1 GAP targets which are
classified with $3.2<\log g<1.9$ and \numax$ < 2.6\times 10^6 \cdot
2^{-\mathit{Kp}}$ in the EPIC is $\sim$\,14\% (corresponding to $\sim 1200$
stars), which is close to the observed fraction. This supports the
conclusions based on the \galaxia\ models that the seismic sample is near
complete.

\section{Conclusion and future outlook}\label{conclusion}
We have analysed all (8,630) stars from the target list of the K2 Galactic
Archaeology Program, and report the detection of oscillations in about 1200
of them; all red giants with  $1.9 \lesssim \log g \lesssim 3.2$.  As a
serendipitous, but valuable by-product of this analysis, we identify a set
of roughly 7500 dwarfs ideally suited for exoplanet galactic exoplanet
occurrence rate studies. 
\begin{itemize} 
\item As expected for the giants, we find that \dnu\ can not be determined
robustly for all the helium-core burning stars despite them showing
oscillations, with a robust measurements of \numax. 
\item However, somewhat unexpectedly we see a discrepancy in star counts between a
galactic synthesis-model prediction and the lower number of observed giants
with a detection of \numax\ $\lesssim 40\,$\muhz, hence affecting many of
the helium-core burning stars.  
Current evidence does not allow us to 
conclude to what degree this is caused by model error, including our ability
to estimate oscillation amplitude and noise in the data, or by seismic
detection bias against helium-core burning giants.  This discrepancy
clearly needs further investigation in future.   
\begin{itemize} 
  \item One would go some way towards addressing this issue by applying the
  current seismic analysis 
  pipelines on \kepler\ data `degraded' to K2 quality; meaning short time
  series and higher noise levels, although not all K2-specific noise
  properties are straightforward to simulate with \kepler\ data.  
  \item Performing a visual inspection along the lines what we show here, but for
  a large randomly selected set of stars, irrespective of pipeline detection
  or not, could indicate how many oscillating giants that where completely
  missed by any of the automated pipeline algorithms. 
  \item It would further help to look at different galactic population synthesis
  tools \citep[e.g. TRILEGAL][]{Girardi05} and to probe into different
  directions of the Galaxy using data from subsequent K2 campaigns, to
  investigate issues with the galactic model predictions.  Already published
  results from later campaigns also indicate reduced effects from spacecraft
  thruster firings, making a comparison with \kepler\ data more informative. 
  \item In addition, spectroscopic $\log g$ values from large unbiased samples of K2
  GAP targets will provide an important independent comparison to reveal
  detection biases in the seismic sample.
\end{itemize}
\item Despite the known biases, it currently appears that the sample of oscillating
giants with $40 \lesssim$ \numax/\muhz\ $\lesssim 270$ and
\numax$ < 2.6\times 10^6 \cdot 2^{-\mathit{Kp}}\,$\muhz\ 
(Figure~\ref{numax-histo}, \ref{numax-kpmag-comb}), 
provide a set of stars with little or no detection
bias, especially if only \numax\ is used (for example in combination with Gaia
parallaxes).  
\item We note that the C1 data does not suitably inform about
detection bias towards luminous giants fainter than about
$\mathit{Kp}\simeq 14.5$.
\item Finally, if users do not take our visual detection confirmation into
account, we suggest to disregard all stars with \numax\ within a few
\muhz\ of the thruster firing frequency if an unbiased set of stars is
required from the C1 seismic data reported here.
\end{itemize}


\acknowledgments
The K2 Galactic Archaeology Program is supported by the National Aeronautics
and Space Administration under Grant NNX16AJ17G issued through the K2 Guest
Observer Program. 
D.S is the recipient of an Australian Research Council Future Fellowship 
(project number FT1400147).
S.M. acknowledges support from NASA grant NNX12AE167 and NNX16AJ17G.
D.H. acknowledges support by the Australian Research Council's Discovery
Projects funding scheme (project number DE140101364) and support by the
National Aeronautics and Space Administration under Grant NNX14AB92G issued
through the Kepler Participating Scientist Program.

\appendix
This appendix is used to show some of the statistical aspects of the
pipelines including biases between pipelines, which are a result of the
different definitions of, and hence methods used to measure, \dnu\ and \numax.
We refer the reader to 
previous comparisons by \citet{Hekker11,Hekker12} that included some of the
pipelines used here, though somewhat earlier and modified versions relative
to what is presented here. 

\begin{figure}
\includegraphics[width=16.6cm]{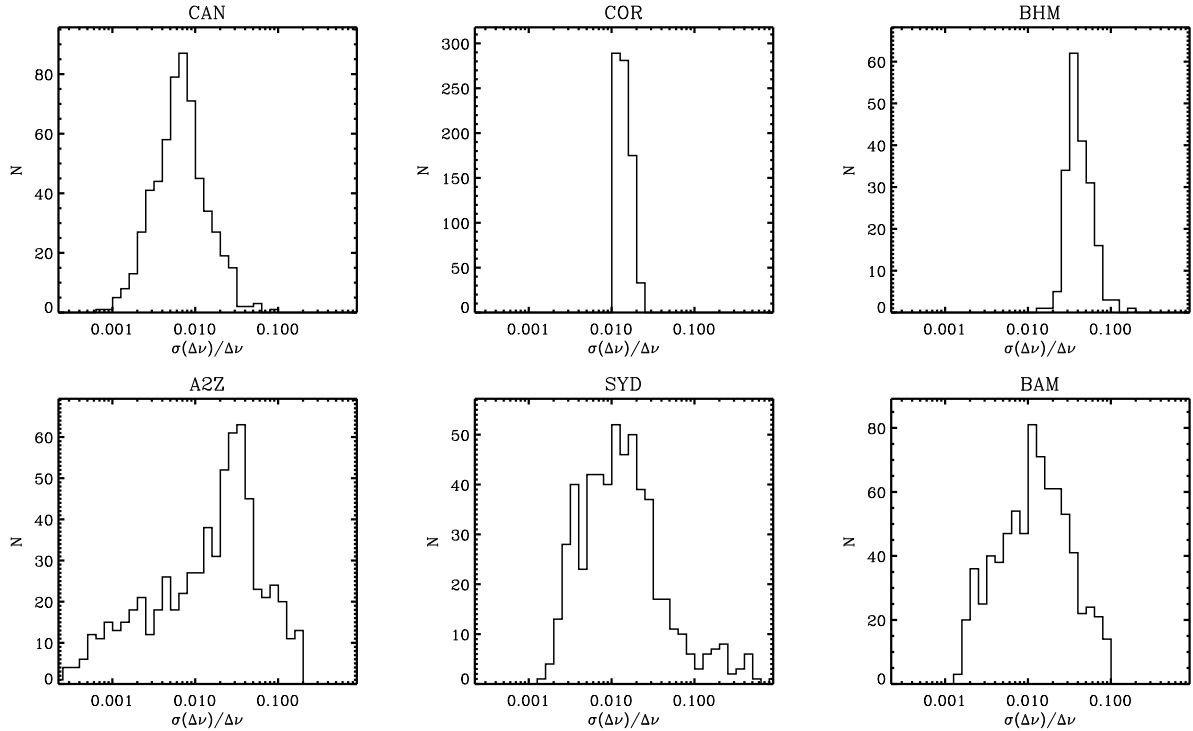}
\caption{Relative uncertainty in \dnu\ reported by each pipeline. The scale on the abscissa is the
  same for all panels.
\label{appfig1}} 
\end{figure}

\begin{figure}
\includegraphics[width=8.8cm]{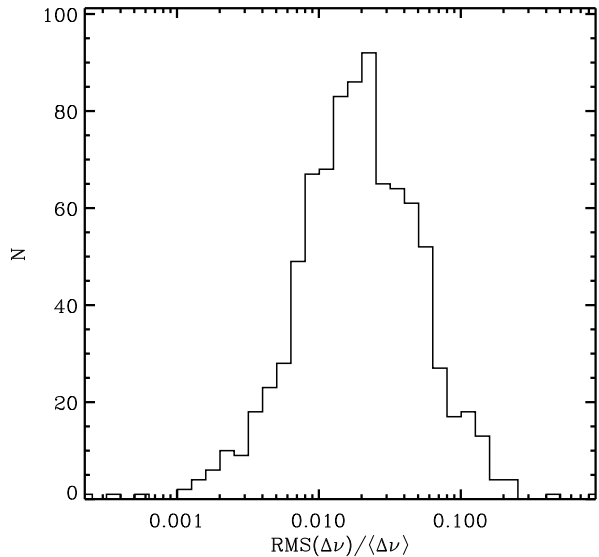}
\caption{Relative RMS scatter of \dnu\ between all pipelines. Only stars
  for which at least two pipelines provided results are shown.
\label{appfig2}} 
\end{figure}

\begin{figure}
\includegraphics[width=16.6cm]{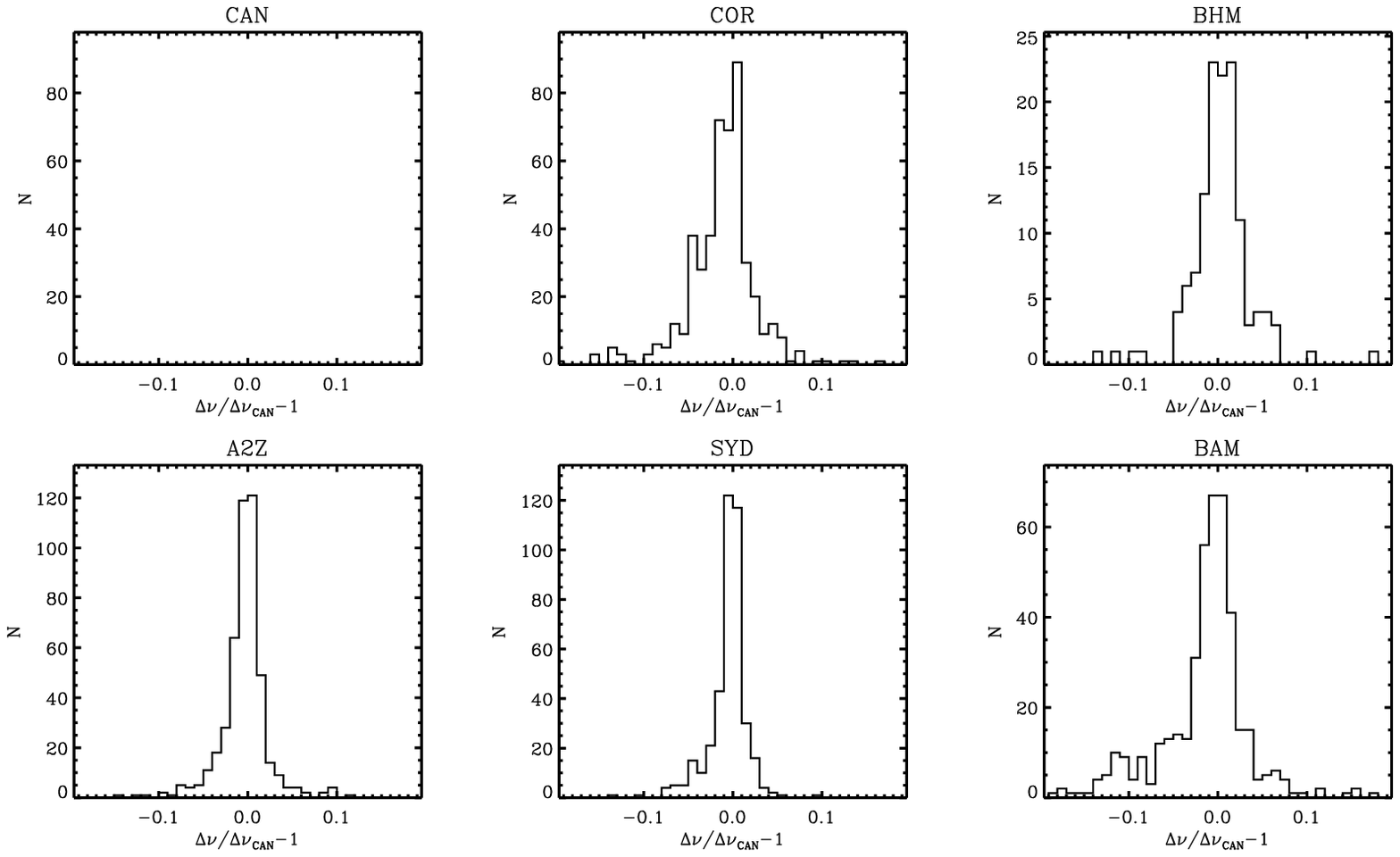}
\caption{\dnu\ relative to the CAN pipeline result. Only stars
  in common with the CAN list are shown. 
\label{appfig3}} 
\end{figure}

\begin{figure}
\includegraphics[width=16.6cm]{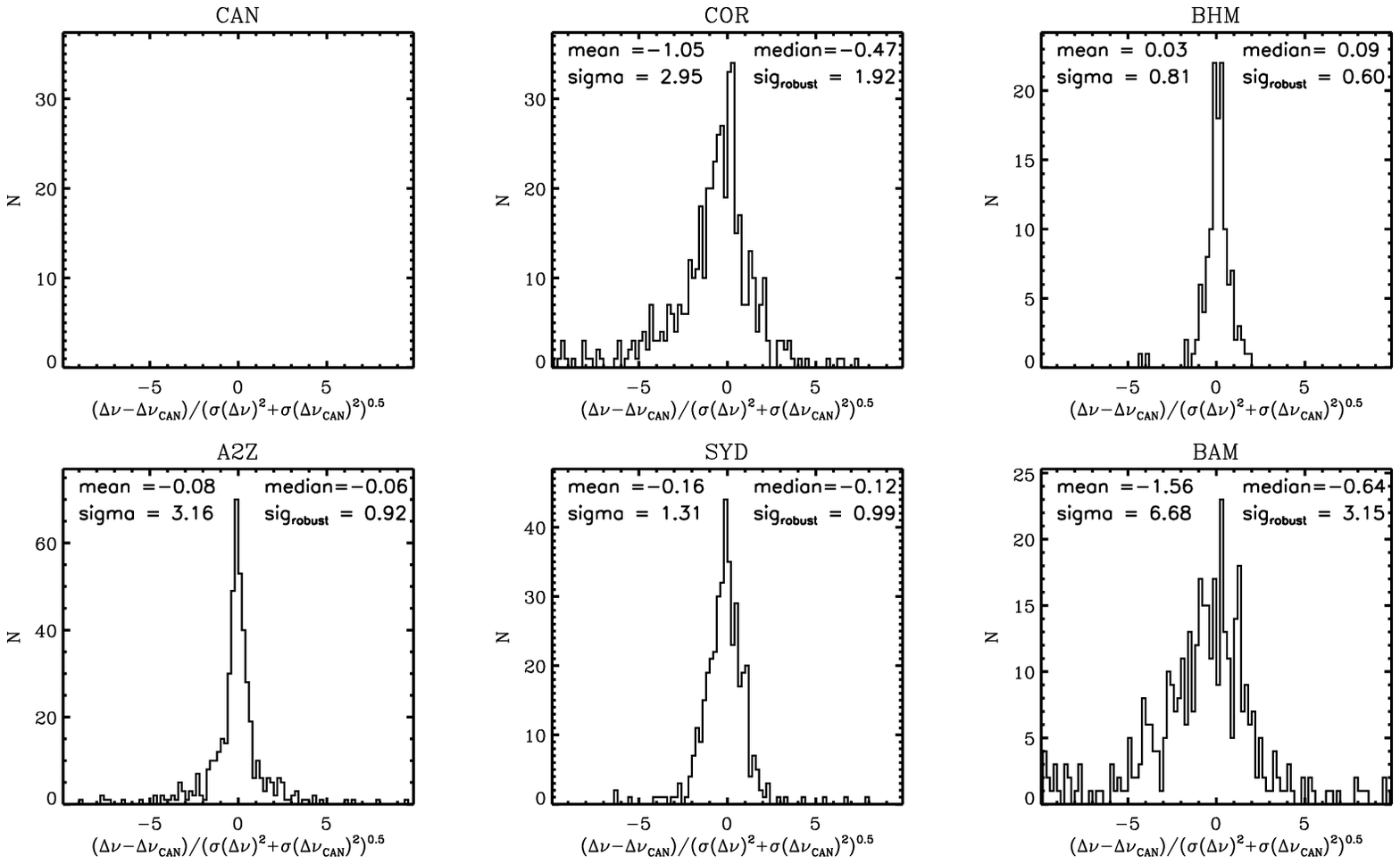}
\caption{\dnu\ deviation from CAN pipeline result relative to quadrature
  uncertainty between the two pipelines. Only stars
  in common with the CAN list are shown.   Straight mean and standard
  deviation is shown in addition to the median and robust standard
  deviation of the distributions.
\label{appfig4}} 
\end{figure}

\begin{figure}
\includegraphics[width=16.6cm]{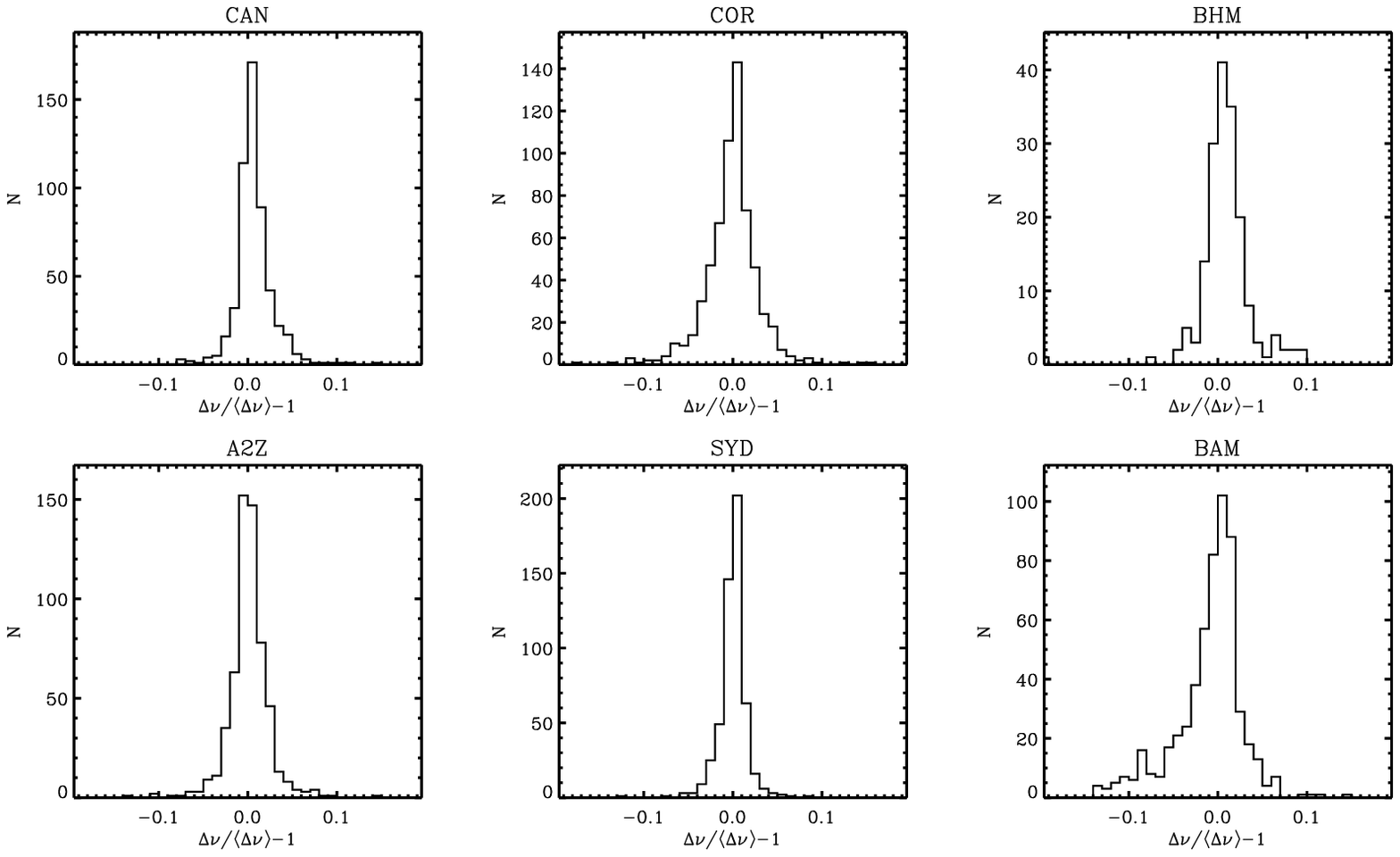}
\caption{\dnu\ relative to the mean across all pipelines. Only stars
  for which at least two other pipelines provided results are shown.
\label{appfig5}} 
\end{figure}

\begin{figure}
\includegraphics[width=16.6cm]{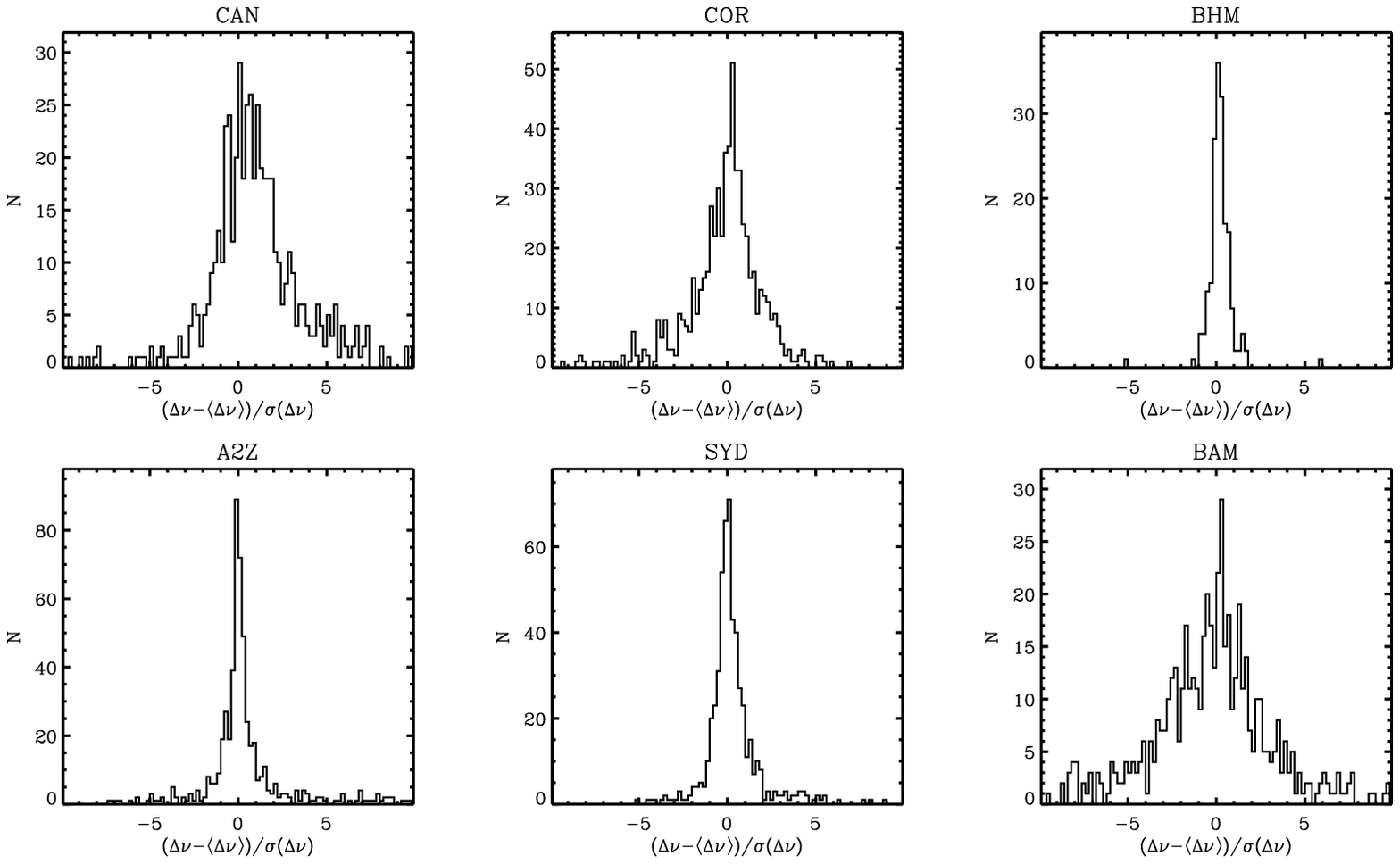}
\caption{\dnu\ deviation from mean across all pipelines relative to
  the uncertainty from the individual pipeline. Only stars for which at least two other pipelines
  provided results are shown. 
\label{appfig6}} 
\end{figure}

We first show results for \dnu.
In Figure~\ref{appfig1} we show the relative uncertainty in \dnu\ reported
by each pipeline. The different methods for deriving uncertainties are
clearly reflected in the different distributions of the relative
uncertainties (see Section~\ref{pipelines}).
In comparison Figure~\ref{appfig2} shows the scatter between the pipelines
for stars in common.  
Systematic pipeline-to-pipeline offsets in \dnu\ are illustrated in Figures~\ref{appfig3}
and \ref{appfig4}, where the reference value is that from CAN, and in Figures~\ref{appfig5}
and \ref{appfig6}, where the reference value is the mean across all pipelines. 

\begin{figure}
\includegraphics[width=16.6cm]{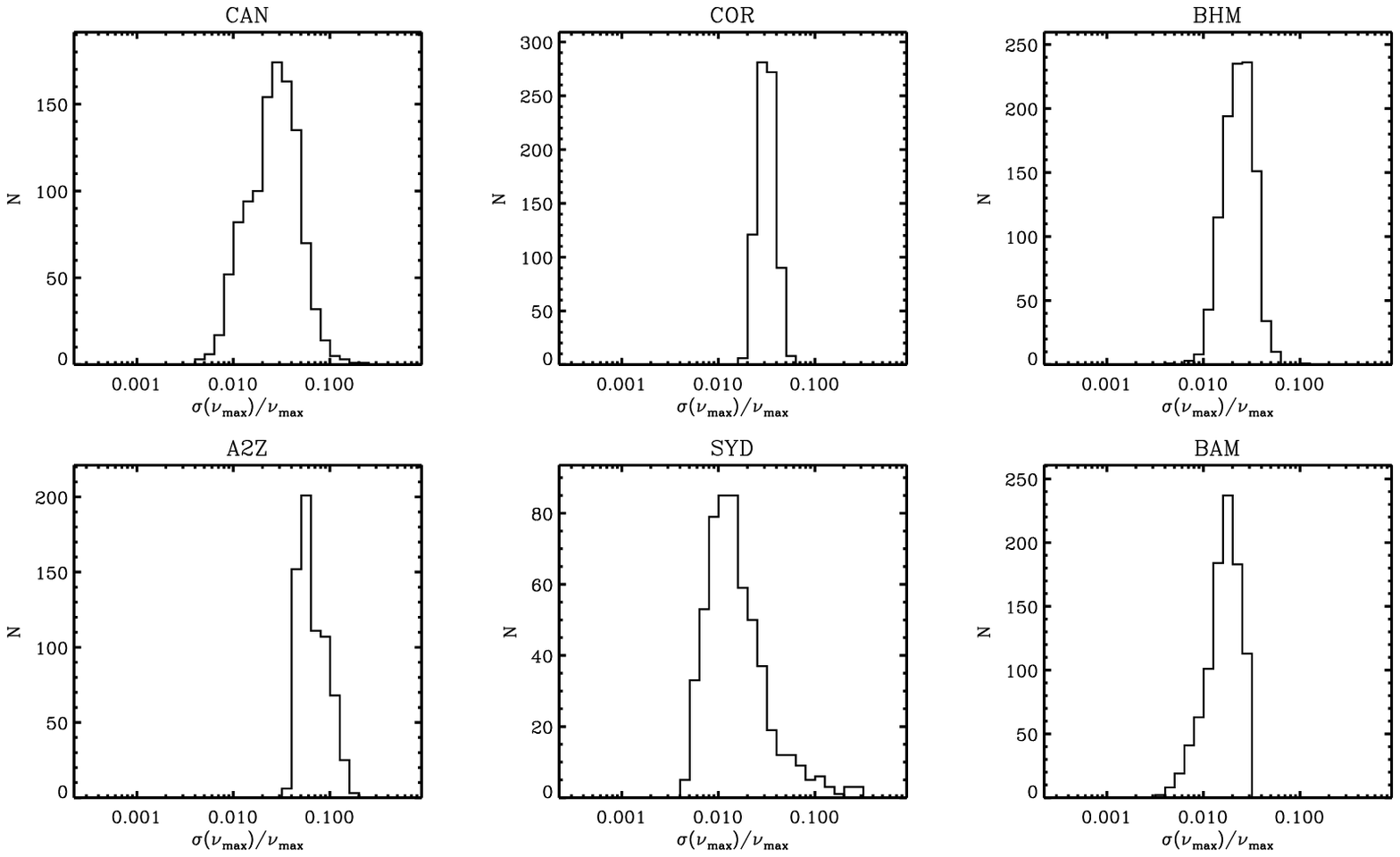}
\caption{Relative uncertainty in \numax.
\label{appfig7}} 
\end{figure}

\begin{figure}
\includegraphics[width=8.8cm]{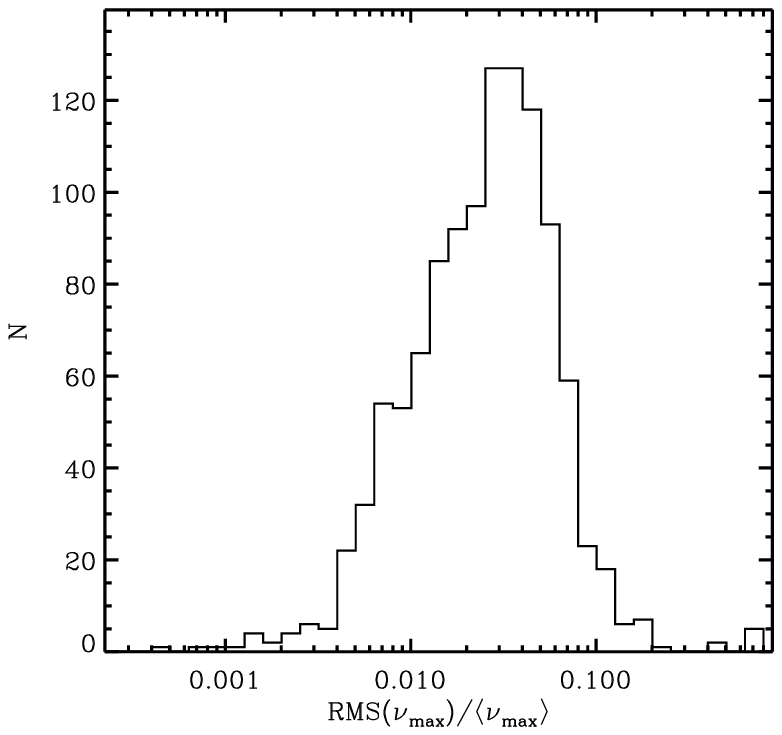}
\caption{Relative RMS scatter of \numax\ between all pipelines. Only stars
  for which at least two pipelines provided results are shown. 
\label{appfig8}} 
\end{figure}

\begin{figure}
\includegraphics[width=16.6cm]{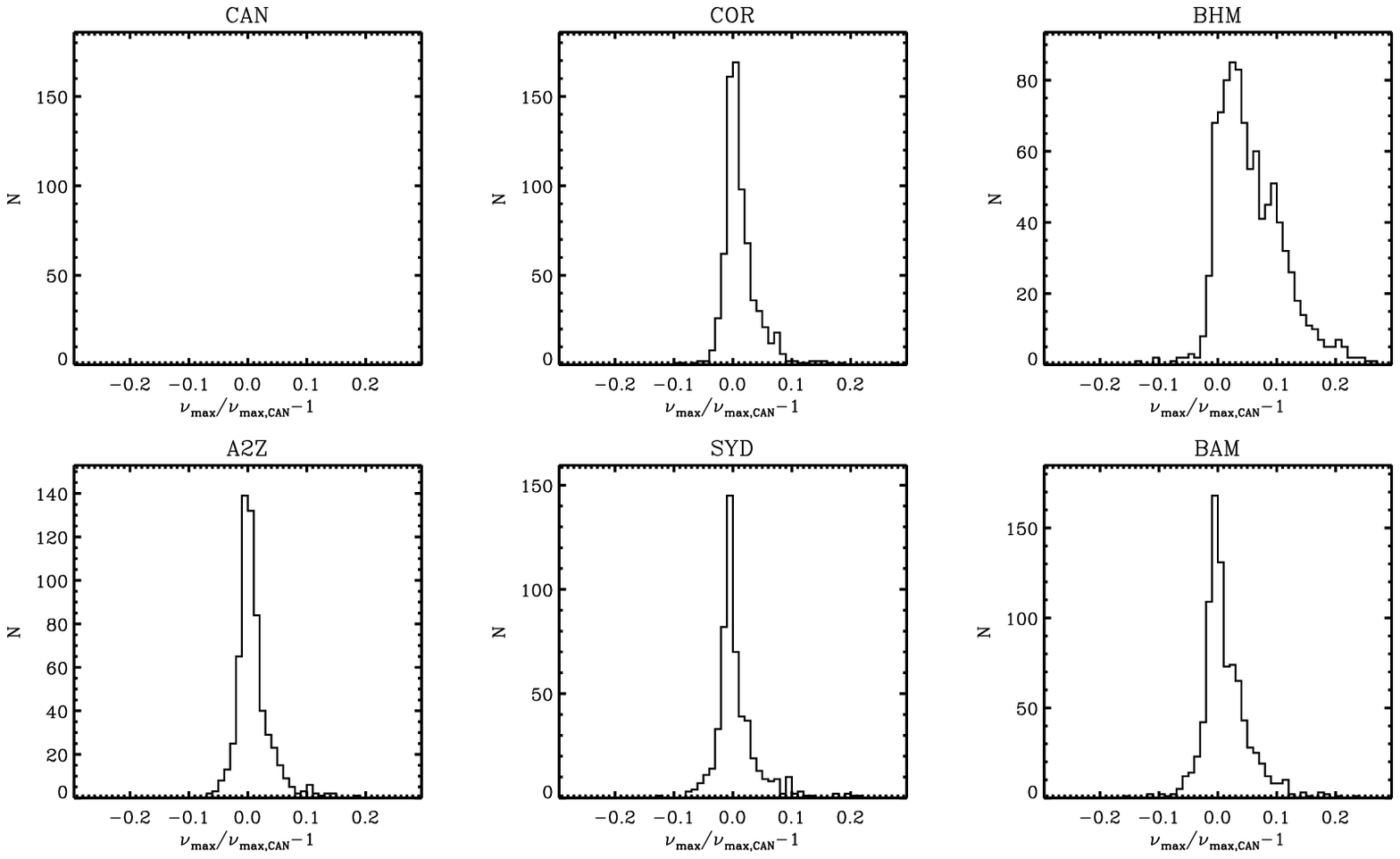}
\caption{\numax\ relative to the CAN pipeline result. Only stars
  in common with the CAN list are shown. 
\label{appfig9}} 
\end{figure}

\begin{figure}
\includegraphics[width=16.6cm]{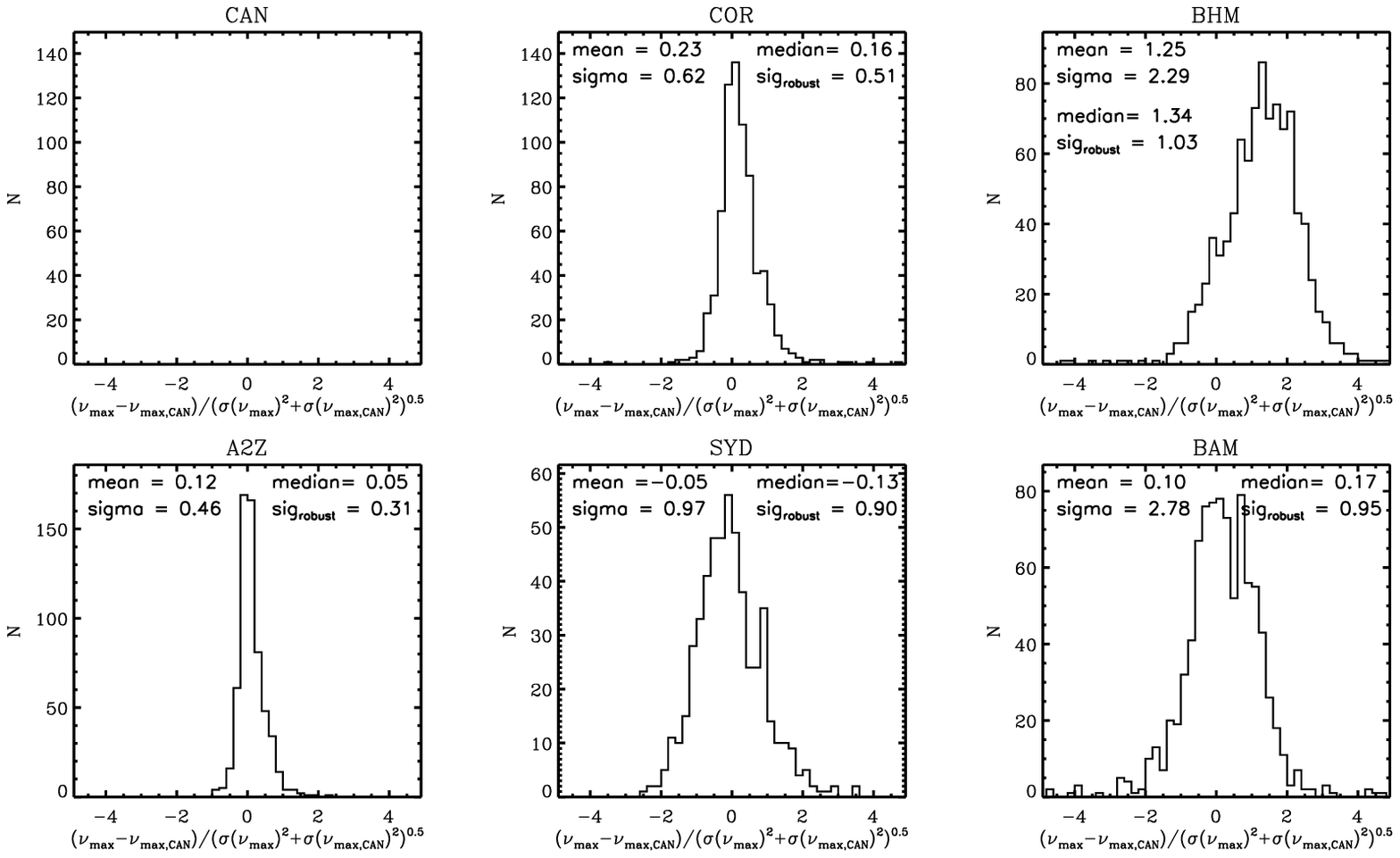}
\caption{\numax\ deviation from CAN pipeline result relative to quadrature
  uncertainty between the two pipelines. Only stars
  in common with the CAN list are shown. Straight mean and standard
  deviation is shown in addition to the median and robust standard
  deviation of the distributions.
\label{appfig10}} 
\end{figure}

\begin{figure}
\includegraphics[width=16.6cm]{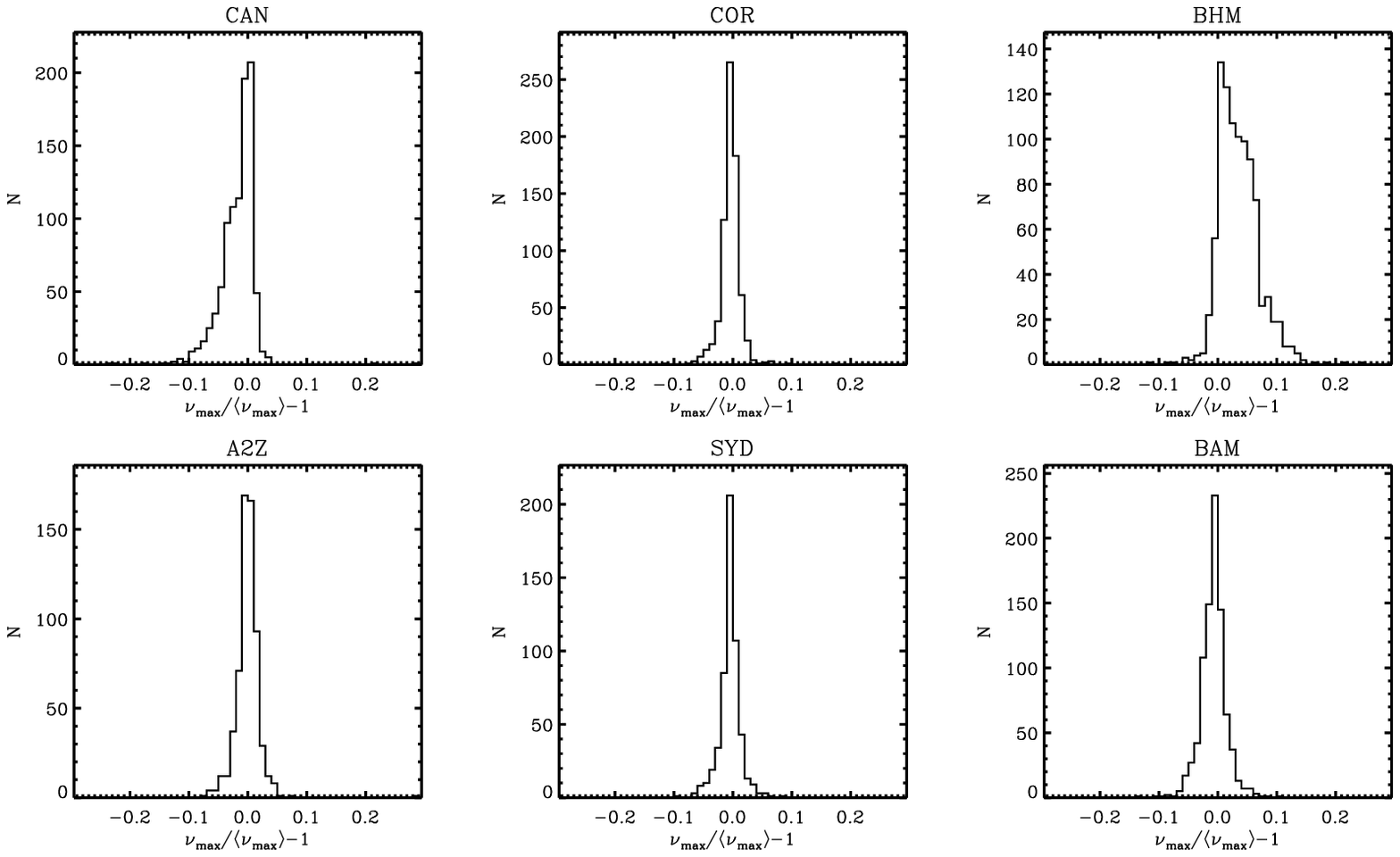}
\caption{\numax\ relative to mean across all pipelines. Only stars
  for which at least two other pipelines provided results are shown.
\label{appfig11}} 
\end{figure}

\begin{figure}
\includegraphics[width=16.6cm]{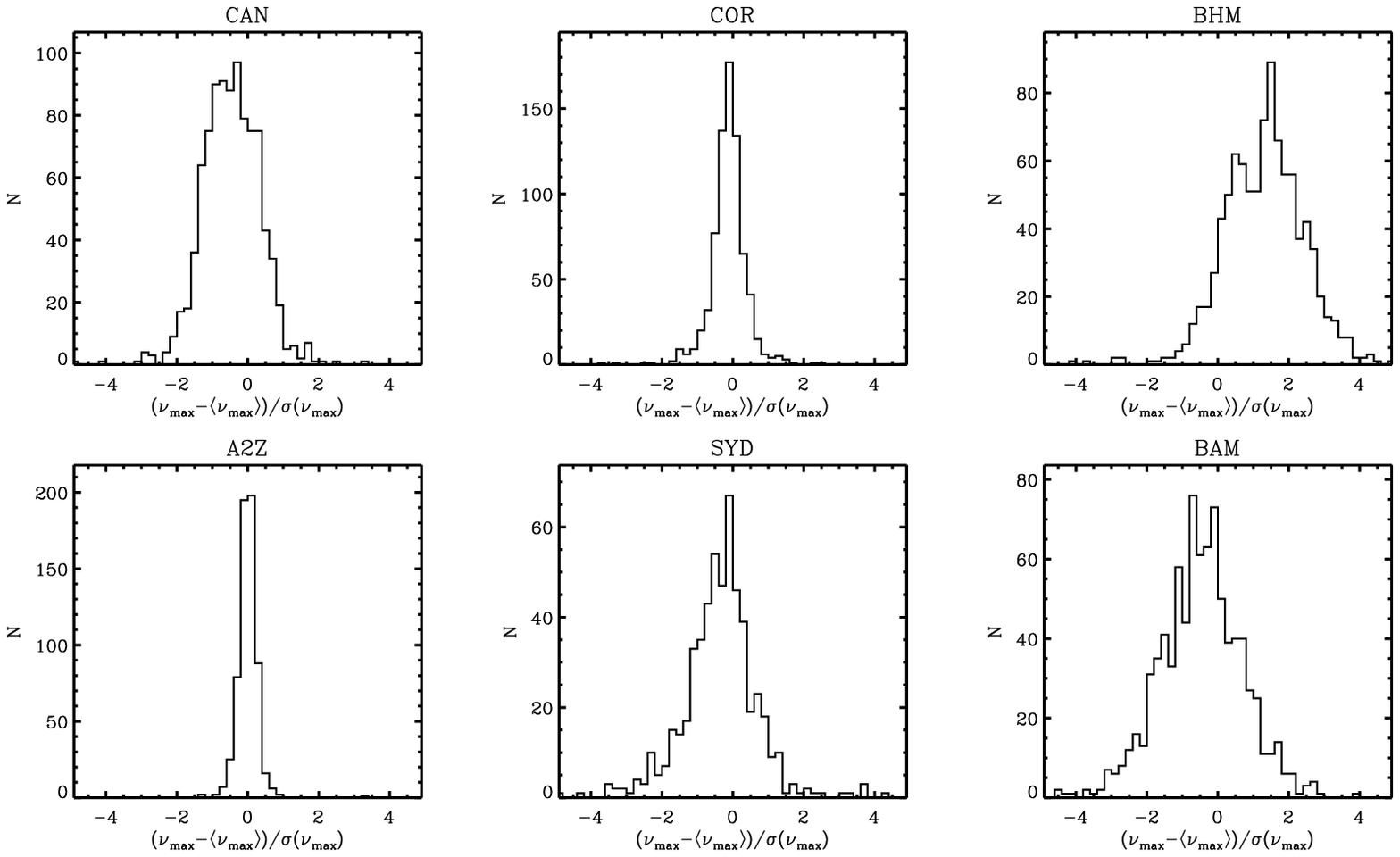}
\caption{\numax\ deviation from mean across all pipelines relative to
  the uncertainty from the individual pipeline. Only stars for which at least two other pipelines
  provided results are shown. 
\label{appfig12}} 
\end{figure}

Now we turn to the \numax\ measurements.
In Figure~\ref{appfig7} we show the relative uncertainty in \numax\ reported
by each pipeline. Like in the case of \dnu, here we also see clear
differences in the distributions of the relative uncertainties.
In comparison Figure~\ref{appfig8} shows the scatter between the pipelines
for stars in common.  
Pipeline-to-pipeline offsets in \numax\ are illustrated in Figures~\ref{appfig9}
and \ref{appfig10}, where CAN is the reference value, and in Figures~\ref{appfig11}
and \ref{appfig12}, where the reference is the mean across all
pipelines.


\end{document}